\documentclass[]{aa}
\usepackage{txfonts}
\usepackage{natbib,twoopt}
\usepackage{amsmath} 
\usepackage[breaklinks=true]{hyperref} 

\bibpunct{(}{)}{;}{a}{}{,} 
\makeatletter
\newcommandtwoopt{\citeads}[3][][]{\href{http://adsabs.harvard.edu/abs/#3}%
{\def\hyper@linkstart##1##2{}%
\let\hyper@linkend\@empty\citealp[#1][#2]{#3}}}
\newcommandtwoopt{\citepads}[3][][]{\href{http://adsabs.harvard.edu/abs/#3}%
{\def\hyper@linkstart##1##2{}%
\let\hyper@linkend\@empty\citep[#1][#2]{#3}}}
\newcommandtwoopt{\citetads}[3][][]{\href{http://adsabs.harvard.edu/abs/#3}%
{\def\hyper@linkstart##1##2{}%
\let\hyper@linkend\@empty\citet[#1][#2]{#3}}}
\newcommandtwoopt{\citeyearads}[3][][]%
{\href{http://adsabs.harvard.edu/abs/#3}
{\def\hyper@linkstart##1##2{}%
\let\hyper@linkend\@empty\citeyear[#1][#2]{#3}}}
\makeatother

\usepackage{color}
\definecolor{mygreen}{RGB}{0,128,0}

\def\pmax{51.5} 
\def\pmaxerr{2.0}
\def\pmaxmin{49.5}
\def\pmaxmax{53.5}

\def\thetamax{92} 

\def\distdraine{1910} 
\def\distdraineerr{80}
\def\distdraineerrpercent{4.2\%}

\def\distrayleigh{1940} 
\def\distrayleigherr{70}
\def\distrayleigherrpercent{3.6\%}

\def\parallax{0.524} 
\def\parallaxerr{0.022}
\begin{document}

\title{The long-period Galactic Cepheid RS~Puppis\thanks{Based on observations made with the NASA/ESA \emph{Hubble Space Telescope}, obtained at the Space Telescope Science Institute, which is operated by the Association of Universities for Research in Astronomy, Inc., under NASA contract NAS 5-26555. These observations are associated with program \#13454.}}
\subtitle{III. A geometric distance from \emph{HST} polarimetric imaging of its light echoes}
\titlerunning{The geometric distance of RS~Pup from \emph{HST} polarimetric imaging of its light echoes}
\authorrunning{P. Kervella et al.}
\author{
P.~Kervella\inst{1}
\and
H.~E.~Bond\inst{2,3}
\and
M.~Cracraft\inst{3}
\and
L.~Szabados\inst{4}
\and
J.~Breitfelder\inst{1,5}
\and
A.~M\'erand\inst{5}
\and
W.~B.~Sparks\inst{3}
\and
A.~Gallenne\inst{6}
\and
D.~Bersier\inst{7}
\and
P.~Fouqu\'e\inst{8}
\and
R.~I.~Anderson\inst{9}
}

\institute{
LESIA, Observatoire de Paris, CNRS, UPMC, Universit\'e Paris-Diderot, PSL Research University, 5 place Jules Janssen, 92195 Meudon, France, \email{pierre.kervella@obspm.fr}.
\and
Department of Astronomy and Astrophysics, 525 Davey Lab., Pennsylvania State University, University Park, PA 16802 USA.
\and
Space Telescope Science Institute, 3700 San Martin Dr., Baltimore, MD 21218 USA.
\and
Konkoly Observatory, H-1525 Budapest XII, P. O. Box 67, Hungary.
\and
European Southern Observatory, Alonso de C\'ordova 3107, Casilla 19001, Santiago 19, Chile.
\and
Universidad de Concepci{\'o}n, Departamento de Astronom\'{\i}a, Casilla 160-C, Concepci{\'o}n, Chile.
\and
Astrophysics Research Institute, Liverpool John Moores University, Liverpool Science Park, 146 Brownlow Hill Liverpool, L3 5RF, United Kingdom.
\and
IRAP, UMR 5277, CNRS, Universit\'e de Toulouse, 14 av. E. Belin, F-31400 Toulouse, France.
\and
Observatoire de Gen\`eve, Universit\'e de Gen\`eve, 51 Ch. des Maillettes, 1290 Sauverny, Switzerland
}
\date{Received ; Accepted}
\abstract
{As one of the most luminous Cepheids in the Milky Way, the 41.5-day RS~Puppis is an analog of the long-period Cepheids used to measure extragalactic distances. An accurate distance to this star would therefore help anchor the zero-point of the bright end of the period-luminosity relation. But, at a distance of about 2\,kpc, RS~Pup is too far away for measuring a direct trigonometric parallax with a precision of a few percent with existing instrumentation.
RS~Pup is unique in being surrounded by a reflection nebula, whose brightness varies as pulses of light from the Cepheid propagate outwards.
We present new polarimetric imaging of the nebula obtained with \emph{HST}/ACS. The derived map of the degree of linear polarization $p_L$ allows us to reconstruct the three-dimensional structure of the dust distribution. To retrieve the scattering angle from the $p_L$ value, we consider two different polarization models, one based on a Milky Way dust mixture and one assuming Rayleigh scattering.
Considering the derived dust distribution in the nebula, we adjust a model of the phase lag of the photometric variations over selected nebular features to retrieve the distance of RS~Pup. We obtain a distance of $\distdraine \pm \distdraineerr$\,pc (\distdraineerrpercent), corresponding to a parallax of $\pi = \parallax \pm \parallaxerr$\,mas. The agreement between the two polarization models we considered is good, but the final uncertainty is dominated by systematics in the adopted model parameters.
The distance we obtain is consistent with existing measurements from the literature, but light echoes provide a distance estimate that is not subject to the same systematic uncertainties as other estimators (e.g. the Baade-Wesselink technique). RS~Pup therefore provides an important fiducial for the calibration of systematic uncertainties of the long-period Cepheid distance scale.
}

\maketitle

\section{Introduction}

Thanks to their very high intrinsic luminosity, long-period Cepheids are observable in distant galaxies and can be used as standard candles up to tens of megaparsecs. As a consequence, Cepheids are a central rung in the extragalactic distance scale \citepads{2010ARA&A..48..673F, 2009ApJS..183..109R, 2011ApJ...730..119R}. The long-period Cepheid \object{RS Puppis} (\object{HD 68860}) is one of the most massive and intrinsically bright stars of its class in the Galaxy. Its physical properties are comparable to the Cepheids observed to measure cosmological distances, which makes it an important fiducial for the calibration of the Cepheid distance scale \citepads[see e.g.~][]{2011ApJ...730..119R, 2001ApJ...553...47F, 2006ApJ...653..843S}.

\object{RS Pup} is unique among Galactic Cepheids in being embedded in a reflection nebula, discovered by \citetads{1961PASP...73...72W}. \citetads{1972A&A....16..252H} demonstrated that features within the nebula show light variations at the Cepheid's pulsation period and argued that these ``echoes'' could be used to derive a geometric distance. \citetads{2008A&A...480..167K} used the EMMI imager on the 3.6-m ESO NTT telescope to monitor the light variations of the nebula and determine phase lags of the brightest knots in the reflection nebula. They determined a very precise geometric distance to RS~Pup of $1992\pm28$~pc. However, \citetads{2009A&A...495..371B} pointed out that this conclusion was based on an assumption that the knots were predominantly located near the plane of the sky; such an assumption is unlikely to be the case because the high efficiency of forward scattering means that the knots will preferentially lie in front of the star.

\citetads{2009A&A...495..371B} did, however, point out that \emph{polarimetric imaging} could be used to determine a purely geometric distance to RS~Pup, because of the fact that the degree of linear polarization of dust-scattered light is maximum for a $\sim$$90^\circ$ scattering angle. This method had originally been proposed by \citetads{1994ApJ...433...19S} as a method for determining distances to extragalactic supernovae, and was one of the arguments for including a polarimetric capability in the Advanced Camera for Surveys (ACS) that was installed in the \emph{Hubble Space Telescope} (\emph{HST}) in 2002. The method was actually first applied by \citetads{2008AJ....135..605S} to the spectacular light echo illuminated by the Galactic luminous transient V838~Monocerotis. The polarimetric distance determined for V838~Mon was verified independently by main-sequence fitting for a sparse open cluster associated with the star \citepads{2007AJ....133..387A}.

In this paper we present new ACS polarimetric imaging of the RS~Pup nebula, from which we obtain a precise geometric distance. Unlike the case of the single outburst of V838~Mon, RS~Pup emits a periodic cycle of light maxima every 41.5~days, leading to a set of nested echoes. These echoes are dominated by dust lying in front of the star. By adopting a polarization scattering model, however, we are nevertheless able to recover the geometric distance.

\section{Observations and data processing\label{observations}}

\subsection{Instrumental setup and raw data reduction\label{instsetup}}

We observed RS~Pup using the Wide Field Channel (WFC) of the ACS onboard the \emph{HST}. We selected the high-throughput F606W filter, which was combined successively with the three polarizers (0, 60 and $120^\circ$). ACS/WFC polarimetry with this filter is now well-calibrated \citepads{2007acs..rept...10C, 2008AJ....135..605S}. This wavelength range is an optimal choice in terms of scattered light intensity, as the high flux from the yellow supergiant combines with the efficient scattering from the dust.
As the field-of-view (hereafter FoV) of the ACS/WFC polarizers (an unvignetted circular FoV with diameter $70''$) is smaller than the $\sim$$3'$ size of the RS Pup nebula, we positioned the Cepheid near the edge of the FoV, and we focused on a bright, high polarization area of the nebula that was identified by \citetads{2012A&A...541A..18K}.
Seven epochs of polarimetric imaging, equally spaced throughout a single pulsation cycle of RS~Pup, were obtained between 2010 March 25 and April 29. We also spent one \emph{HST} orbit to obtain a two-filter (F435W and F606W) unpolarized image of the entire nebula in the full WFC field, in order to map the faint extensions of the nebula and accurately estimate the sky background. Finally we also obtained two observations of nearby bright Point Spread Function (PSF) stars of two different colors matching the color range of the Cepheid. The complete list of the exposures obtained in our 10-orbit \emph{HST} program is presented in Table~\ref{hst_log}.

\onltab{ 
\begin{table*}
\caption{Log of the \emph{HST} observations of RS~Pup and its associated PSF calibrators.}
\label{hst_log}
\begin{tabular}{llllcclll}
\hline \hline \noalign{\smallskip}                 
Dataset & Target & UT Date & UT Time  & JD$_\odot$ & Exp. & Aperture & Filter & Polarizer \\
 &  &    &  & $-$ 2\,450\,000  & (s) &  &  &  \\
\hline \noalign{\smallskip}                 
JB7301010 & RS Pup & 2010 March 25 & 12:50:46 & 5281.038 & 468 & WFC1 & F606W & POL0V \\
JB7301020 & RS Pup & 2010 March 25 & 12:58:16 & 5281.043 & 468 & WFC1 & F606W & POL60V \\
JB7301030 & RS Pup & 2010 March 25 & 13:05:46 & 5281.048 & 468 & WFC1 & F606W & POL120V \\
JB7309010 & PSF-Red & 2010 March 25 & 14:31:49 & 5281.108 & 468 & WFC1 & F606W & POL0V \\
JB7309020 & PSF-Red & 2010 March 25 & 14:39:19 & 5281.113 & 468 & WFC1 & F606W & POL60V \\
JB7309030 & PSF-Red & 2010 March 25 & 14:46:49 & 5281.118 & 468 & WFC1 & F606W & POL120V \\
\hline \noalign{\smallskip}
JB7310010 & RS Pup & 2010 March 26 & 06:27:26 & 5281.771 & 1294 & WFCENTER & F435W &  \\
JB7310020 & RS Pup & 2010 March 26 & 06:39:30 & 5281.780 & 800 & WFCENTER & F606W &  \\
\hline \noalign{\smallskip}
JB7302010 & RS Pup & 2010 March 31 & 07:54:07 & 5286.831 & 468 & WFC1 & F606W & POL0V \\
JB7302020 & RS Pup & 2010 March 31 & 08:01:38 & 5286.837 & 468 & WFC1 & F606W & POL60V \\
JB7302030 & RS Pup & 2010 March 31 & 08:09:07 & 5286.842 & 468 & WFC1 & F606W & POL120V \\
\hline \noalign{\smallskip}
JB7303010 & RS Pup & 2010 April 6 & 05:32:44 & 5292.733 & 468 & WFC1 & F606W & POL0V \\
JB7303020 & RS Pup & 2010 April 6 & 06:09:50 & 5292.759 & 468 & WFC1 & F606W & POL60V \\
JB7303030 & RS Pup & 2010 April 6 & 06:17:20 & 5292.764 & 468 & WFC1 & F606W & POL120V \\
\hline \noalign{\smallskip}
JB7308010 & PSF-Blue & 2010 April 6 & 07:48:45 & 5292.827 & 470 & WFC1 & F606W & POL0V \\
JB7308020 & PSF-Blue & 2010 April 6 & 07:56:15 & 5292.833 & 470 & WFC1 & F606W & POL60V \\
JB7308030 & PSF-Blue & 2010 April 6 & 08:03:47 & 5292.838 & 470 & WFC1 & F606W & POL120V \\
\hline \noalign{\smallskip}
JB7304010 & RS Pup & 2010 April 11 & 04:29:10 & 5297.689 & 468 & WFC1 & F606W & POL0V \\
JB7304020 & RS Pup & 2010 April 11 & 05:07:09 & 5297.715 & 468 & WFC1 & F606W & POL60V \\
JB7304030 & RS Pup & 2010 April 11 & 05:14:39 & 5297.720 & 468 & WFC1 & F606W & POL120V \\
\hline \noalign{\smallskip}
JB7305010 & RS Pup & 2010 April 18 & 08:58:54 & 5304.875 & 468 & WFC1 & F606W & POL0V \\
JB7305020 & RS Pup & 2010 April 18 & 09:06:24 & 5304.881 & 468 & WFC1 & F606W & POL60V \\
JB7305030 & RS Pup & 2010 April 18 & 09:13:54 & 5304.886 & 468 & WFC1 & F606W & POL120V \\
\hline \noalign{\smallskip}
JB7306010 & RS Pup & 2010 April 24 & 05:33:09 & 5310.732 & 468 & WFC1 & F606W & POL0V \\
JB7306020 & RS Pup & 2010 April 24 & 05:40:39 & 5310.737 & 468 & WFC1 & F606W & POL60V \\
JB7306030 & RS Pup & 2010 April 24 & 05:48:09 & 5310.743 & 468 & WFC1 & F606W & POL120V \\
\hline \noalign{\smallskip}
JB7307010 & RS Pup & 2010 April 29 & 22:36:21 & 5316.442 & 468 & WFC1 & F606W & POL0V \\
JB7307020 & RS Pup & 2010 April 29 & 23:29:30 & 5316.479 & 468 & WFC1 & F606W & POL60V \\
JB7307030 & RS Pup & 2010 April 29 & 23:37:00 & 5316.485 & 468 & WFC1 & F606W & POL120V \\
\hline                 
\end{tabular}
\tablefoot{JD$_\odot$ is the heliocentric Julian date. The values listed in the `JD$_\odot$' and `UT time' columns correspond to mid-exposure. `Exp.' is the individual exposure time in seconds.}
\end{table*}
} 

The raw images were processed following the procedure described in \citetads{2008AJ....135..605S}.
The data reduction started by retrieving the {\tt flt} files, which are flat-fielded but not distortion corrected, from the MAST archive.
There were two flat fielded files for each visit/filter/polarizer combination. Each set of two {\tt flt} files were processed using the {\tt astrodrizzle}\footnote{\url{http://www.stsci.edu/hst/HST_overview/drizzlepac}} software \citepads{2012drzp.book.....G} that performed distortion corrections, removed cosmic rays, and shifted and aligned all of the images of RS~Pup. In some cases, the task {\tt tweakreg} was used to ensure better alignment of the images. All of the visits were set to align with the POL0 observation of the first \emph{HST} visit. In order not to affect any of the polarizer parameters such as polarization angle or degree, the sky background was not subtracted from the images, and there was no rotation applied during this step. The images were drizzled to the native ACS/WFC scale of 0.05\,arcsec/pixel. The result of this processing was a single image for each polarizer (POL0V, POL60V and POL120V) for each of the visits (Table~\ref{hst_log}).

\subsection{PSF wings and sky background subtraction\label{psfsubtraction}}

The subtraction of the stray light from the very bright central star and the sky background is an important step of the processing, as their contributions affect the derived polarimetric quantities. To estimate the stray light from RS Pup, we considered the observations of the two PSF reference stars, \object{HD 71670} (hereafter PSF-Red) and \object{HD 68111} (PSF-Blue), of similar apparent brightness as RS Pup (Table~\ref{psfcalibrators}).

\begin{table}
\caption{Properties of the PSF calibrators of RS Pup.}
\label{psfcalibrators}
\begin{tabular}{llccc}
\hline \hline \noalign{\smallskip}
HD & Type & $m_B$ & $m_V$ & $m_R$ \\     
\hline \noalign{\smallskip}
68111 & G3\,Ib & $9.30 \pm 0.02$ & $8.183 \pm 0.011$ & $7.60 \pm 0.02$ \\
71670 & K5\,III & $9.26 \pm 0.02$ & $7.593 \pm 0.011$ & $6.50 \pm 0.02$ \\
\hline
\end{tabular}
\tablefoot{The magnitudes were taken from \citetads{2010PASP..122.1437P}.}
\end{table}

We created a reference PSF image for each polarizing filter POL0V, POL60V and POL120V.
We then masked the background stars and artefacts present in the PSF images using a combination of our two reference stars. The resulting combined PSF image (that is devoid of background stars) is affected by a high noise level, as well as the combination of the artefacts present in the individual images.
Directly subtracting these images from RS~Pup's frames results in a degraded SNR, particularly in the low flux parts of the images. To mitigate this effect, we computed the radial median profile of this PSF image over concentric rings. Based on this smooth median profile, we generated a synthetic PSF image that presents a clean and better estimated background level.
%
Residual ghost images of the telescope pupil are present in the RS~Pup images and not in the PSF composite images. Their position and morphology is complex and variable, but they are easily identified and are not used in the polarization analysis.

\begin{figure}[]
\centering
\includegraphics[width=4.4cm]{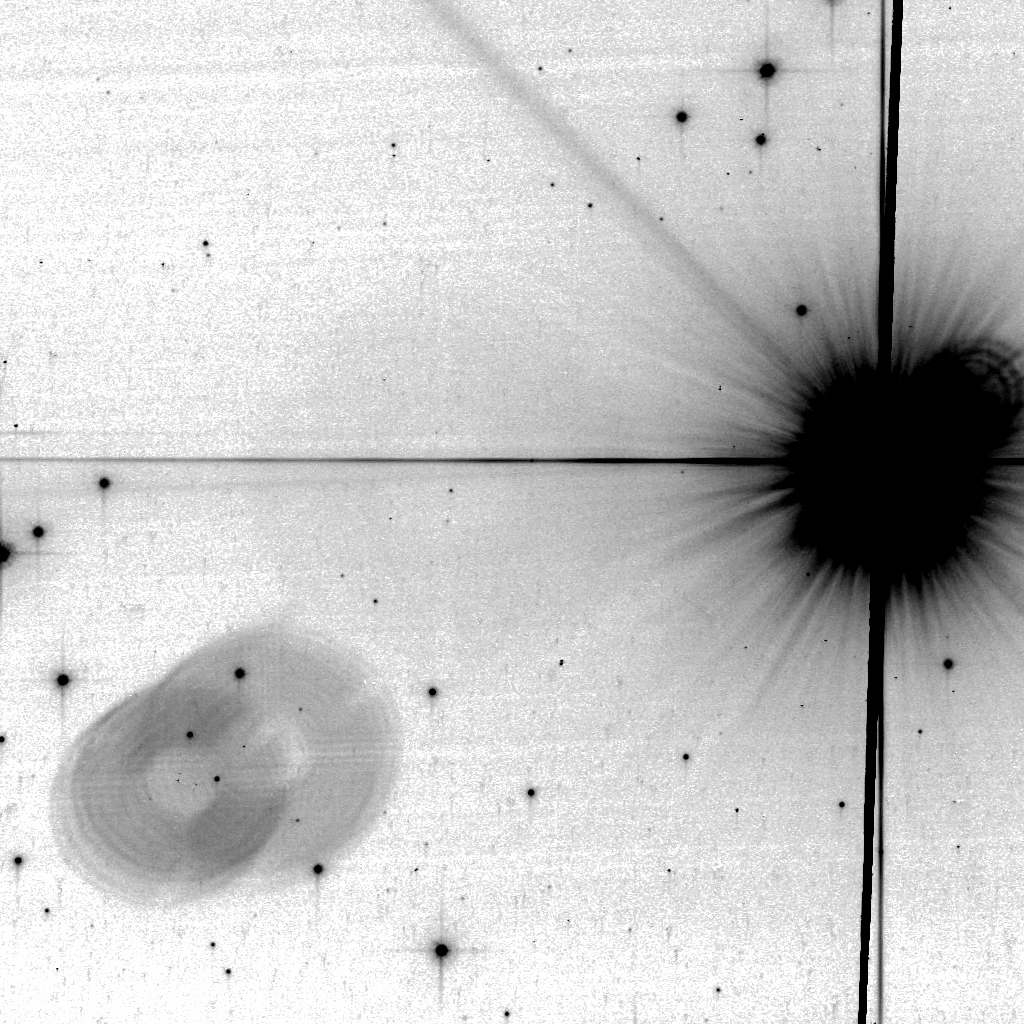} 
\includegraphics[width=4.4cm]{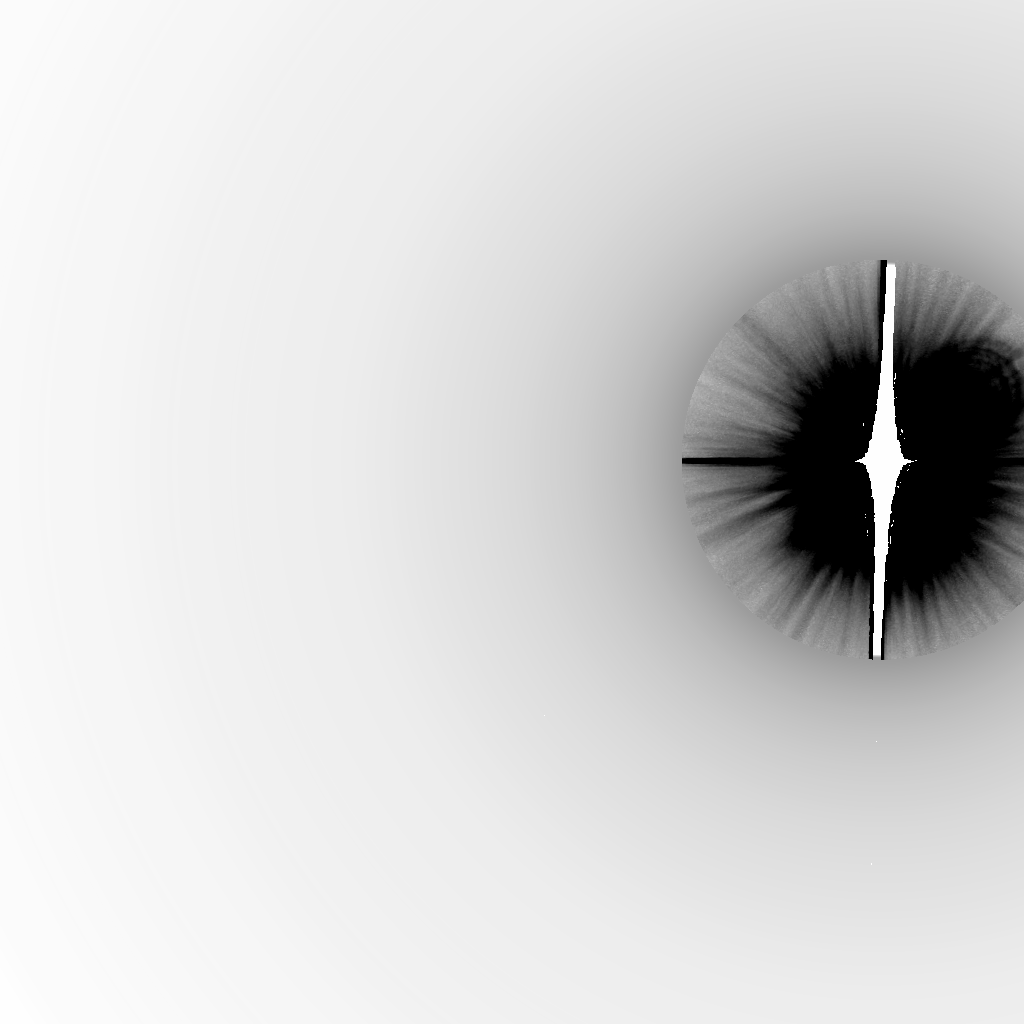} 
\caption{Image of the PSF star HD\,68111 (PSF-Blue) in the F606W+POL0V polarizing filter setup (left) and the derived composite synthetic PSF (right).\label{psfcomposite}}
\end{figure}

The sky background of the images of RS~Pup is difficult to estimate, as the light-scattering, variable surface brightness circumstellar nebula covers most of the WFC1 aperture. We however identified two sufficiently extended regions that present the lowest background level. We checked that the average flux level in these areas is consistent with the sky background measured at larger distances from RS~Pup, using the wide field image obtained on 2010 March 26 in the F606W filter (WFCENTER aperture setup). We subtracted this constant background uniformly from the polarized images of RS~Pup.

\begin{figure}[]
\centering
\includegraphics[width=4.4cm]{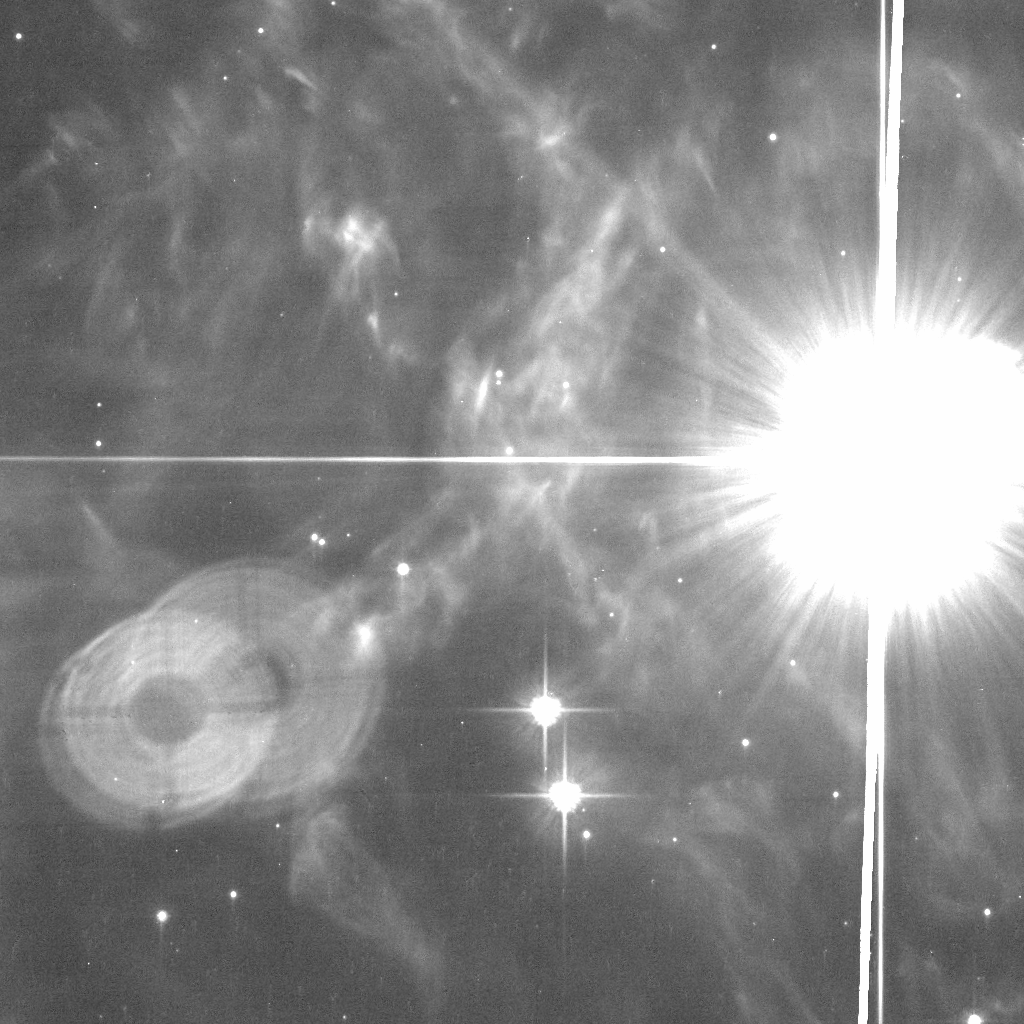} 
\includegraphics[width=4.4cm]{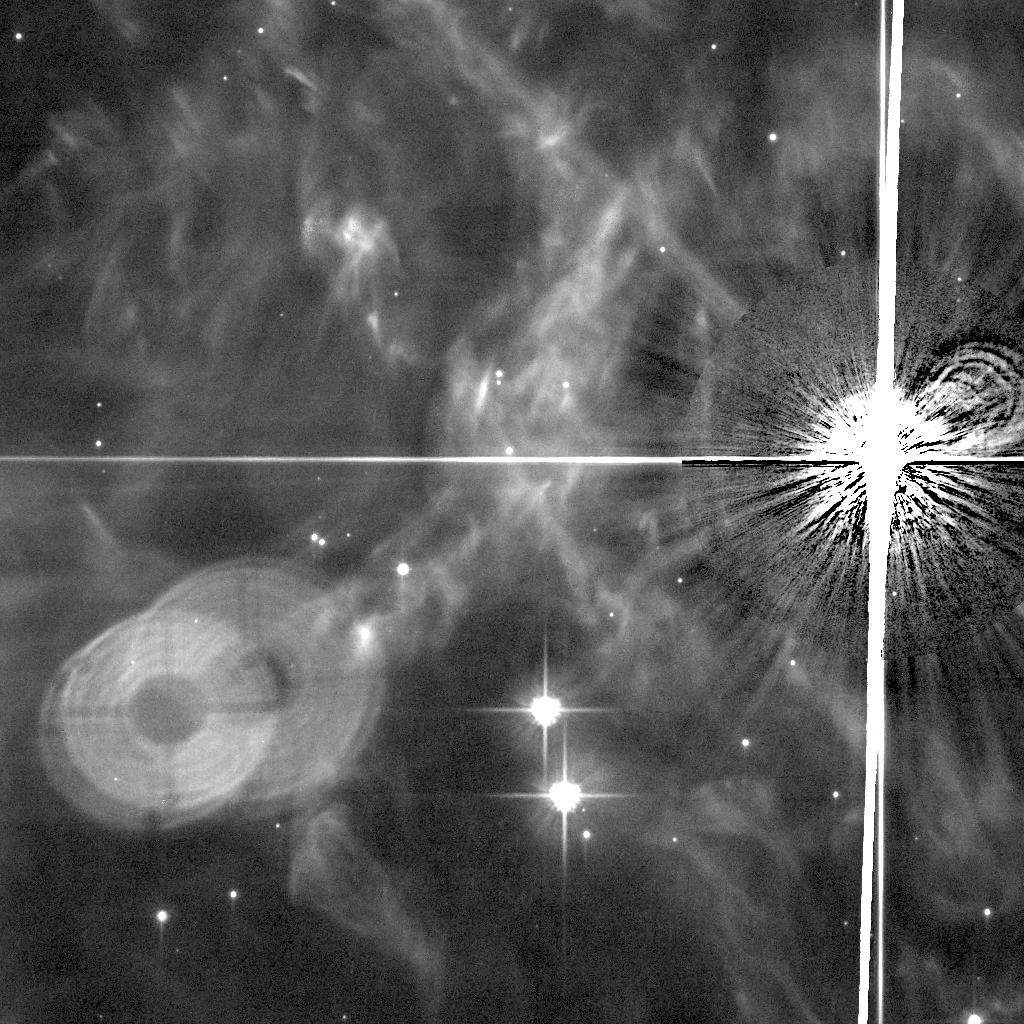} 
\caption{First observing epoch of RS~Pup before (left) and after (right) the subtraction of the PSF wings and sky background, with the same grey scale (proportional to the square root of the flux count). The ``double doughnut'' feature in the lower left quadrant of the image is an optical artefact. \label{psfsubtractionFig}}
\end{figure}

The flux normalization factor of the PSF star to the changing brightness of RS~Pup was computed using the synthetic light curve of the Cepheid in the F606W filter, for each \emph{HST} observing phase.
This light curve of RS~Pup was derived from a combined fit of multicolor photometry \citepads[see][for details]{paper4}, and considering a magnitude $m(F606W) = 7.67 \pm 0.01$ for the PSF-Blue star (HD\,68111).
The flux-normalized PSF images were subtracted from the RS~Pup images, for each polarization filter separately. An example of the result of this subtraction is presented in Fig.~\ref{psfsubtractionFig}. Residual artefacts are present in the subtracted image. They include in particular diffraction and saturation spikes, as well as several pupil ghosts, but their angular extent is limited and they are located outside of the regions of interest (see Sect.~\ref{selection}).

\subsection{Polarimetric quantities}

The Stokes parameters $I$, $Q$ and $U$, the degree of linear polarization $p_L$ and the polarization electric-vector position angle in detector coordinates  $\theta_D$ were derived from the PSF subtracted images of RS~Pup using the approach described by \citetads{2008AJ....135..605S} and \citetads{1999PASP..111.1298S}. The expression of these quantities are presented in Appendix~\ref{appendixpolar}. We adopted the position angles of the electric vectors of the polarizers and the photometric throughputs calibrated by \citetads{2004acs..rept....9B}. The $p_L$ values were debiased considering the photometric signal-to-noise ratio (SNR) using the bias correction listed in Table~A.1 of \citetads{1999PASP..111.1298S}.
The uncertainties were propagated to the Stokes parameters starting from the photometric error bars produced by the ACS processing pipeline.
The $p_L$ and $\theta_D$ values were computed only where the photometric SNR in the unpolarized flux image is sufficiently high ($I/\sigma(I) > 5$), to avoid a divergence in the normalization by $I$. As the nebula is illuminated by the Cepheid (single source), the polarization electric-vector position angle $\theta_D$ is always perpendicular to the dust-star direction over the whole surface of the nebula, as already observed by \citetads{2012A&A...541A..18K}.

\begin{figure*}[]
\centering
\includegraphics[width=9cm]{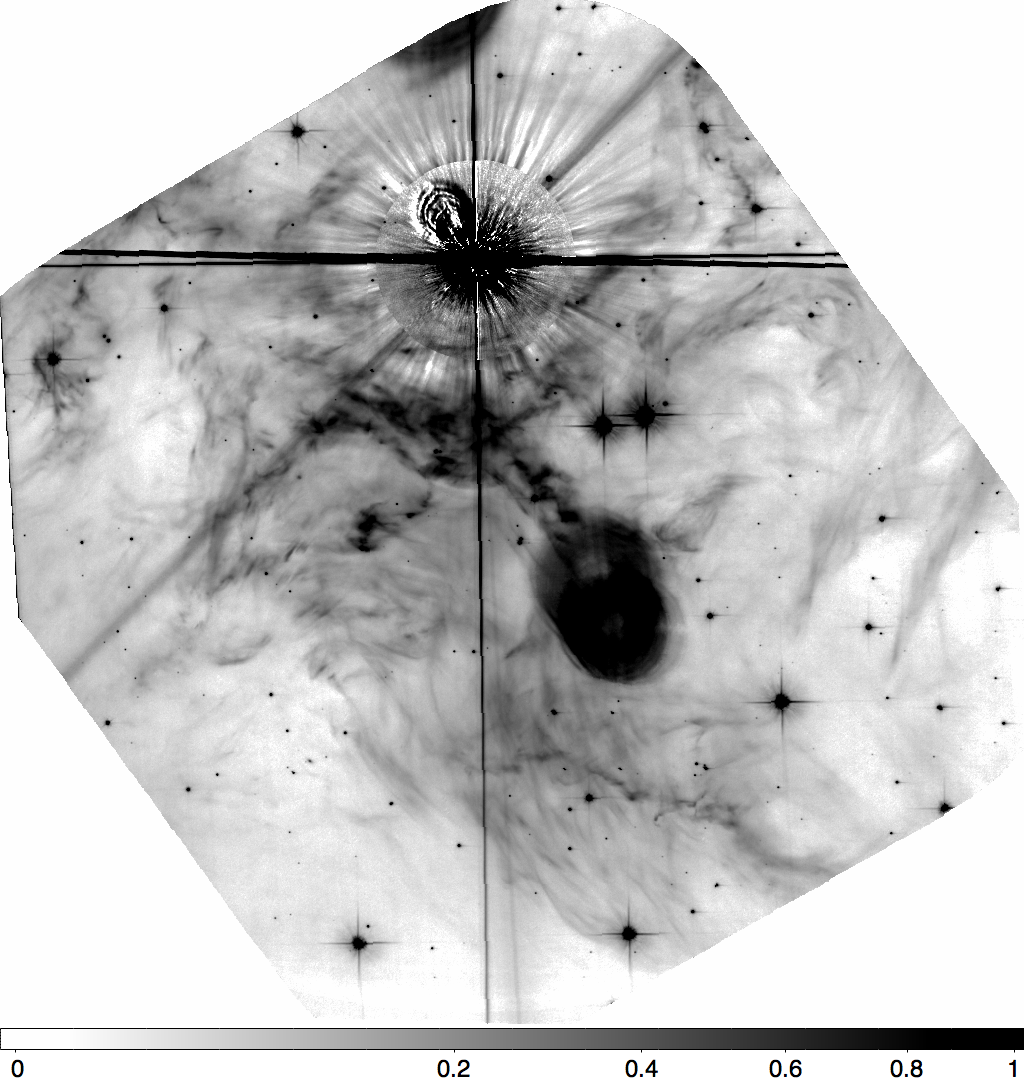} \hspace{1mm} 
\includegraphics[width=9cm]{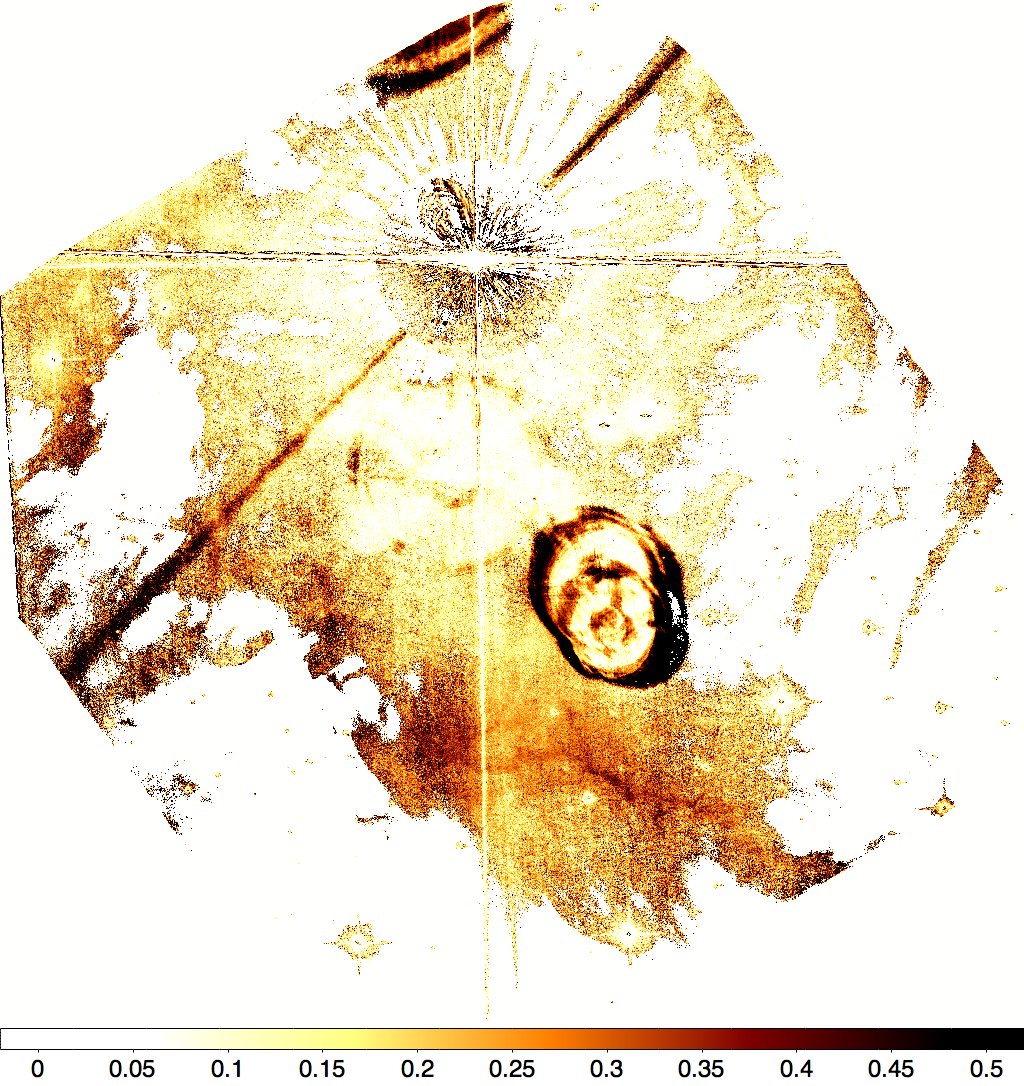} 
\caption{Average maps of the Stokes $I$ (left, in e$^-$\ s$^{-1}$) and degree of linear polarization $p_L$ (right), computed over the seven observing epochs of RS~Pup (North is up and East to the left).\label{IandPL}}
\end{figure*}

\subsection{Phase map and photometry of the nebula}

\subsubsection{Period and reference epoch}

For matching correctly the phases of the light echoes occurring in the nebulosity around RS~Pup, knowledge of the accurate value of the pulsation period of the Cepheid is essential. Therefore, the pulsation period and its variations have been determined with the method of the $O-C$ diagram \citepads{2005ASPC..335....3S}. A discussion of the determination of the period of RS~Pup for the epochs of our \emph{HST}/ACS observations will be presented in \citetads{paper4} (hereafter Paper~IV). We adopt a period $P=41.5117$\,days for the epoch of our ACS observations (2010).

\begin{figure}[]
\centering
\includegraphics[width=\hsize]{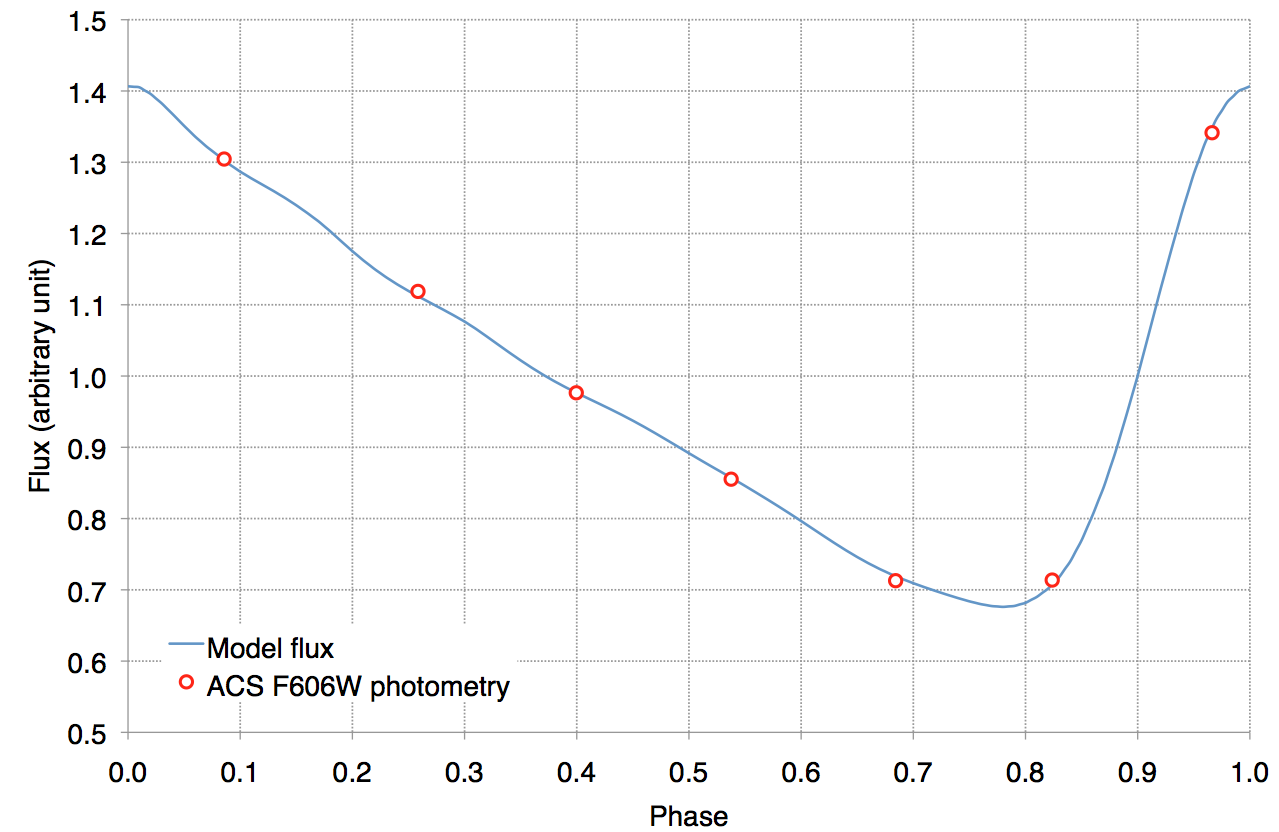}
\caption{Synthetic light curve of RS~Pup (blue curve) in the F606W filter with the ACS photometry derived from the ghost image of the telescope pupil (open circles) phased with $T_0 = 2\,455\,252.64$ (JD$_\odot$) and $P=41.5117$\,days.\label{RSPup-F606W}}
\end{figure}

A simple fit of the light curve to our ACS photometry is the most direct way to obtain a high precision date for the maximum light relevant for our analysis, as we are insensitive to possible random period variations \citepads{2009AstL...35..406B}. This approach also presents the advantage to avoid any shift in the maximum light epoch between the F606W filter and the ephemeris available in the literature (usually in the $V$ or $B$ bands). Unfortunately, the central region close to RS~Pup is heavily saturated in the ACS images, and it is particularly difficult to directly obtain accurate photometry of the star. But the ghost image of the telescope pupil that is present south-southwest of the star in Fig.~\ref{IandPL} (left panel) provides an unsaturated proxy to the stellar flux. Although the equivalent photometric transmission of this pupil image is unknown, its stability is sufficient to derive accurate relative fluxes from our seven observing epochs. Through a fit of the synthetic F606W light curve of RS~Pup to the measured relative photometry of the pupil ghost image, we obtain an heliocentric Julian date $T_0 = 2\,455\,252.64 \pm 0.07$ for the maximum light (Fig.~\ref{RSPup-F606W}). It should be noted that \citetads{2014A&A...566L..10A} demonstrated that the pulsation of RS~Pup shows some degree of cycle-to-cycle variability in radial velocity amplitude and phase, but at a level that does not affect the present light echo analysis.

\subsubsection{Phase offsets over the nebula \label{3paramFit}}

The propagation of the light echoes in the nebula of RS\,Pup are clearly visible in the sequence of our seven epochs of ACS imaging\footnote{An animated video sequence showing the propagation of the light echoes of RS\,Pup can be downloaded from \url{http://hubblesite.org/newscenter/archive/releases/star/2013/51/}}.
We adopt a two-step approach to determine the phase offset of the nebular features. Following \citetads{2008A&A...480..167K},
we first adjust the synthetic light curve of RS~Pup in the F606W filter normalized to unit amplitude and zero average $f(\phi)$ using a 3-parameter model:
\begin{equation}
\label{photomModel}
f(\alpha,\delta,\phi_i) = A\ f(\phi_i - \Delta \phi) + f_0
\end{equation}
where $\phi_i$ is the phase of the considered observation $i$ (i.e. the phase of the Cepheid itself at the corresponding epoch), $A$ is the amplitude of the photometric variation at coordinates $(\alpha,\delta)$ in the nebula, $f_0$ the average photometric flux, and $\Delta \phi$ the phase offset relative to the Cepheid's photometric variation. For each position over the nebula, we therefore derive a triplet $(A, f_0, \Delta \phi)$.

\begin{figure*}[]
\centering
\includegraphics[width=6cm]{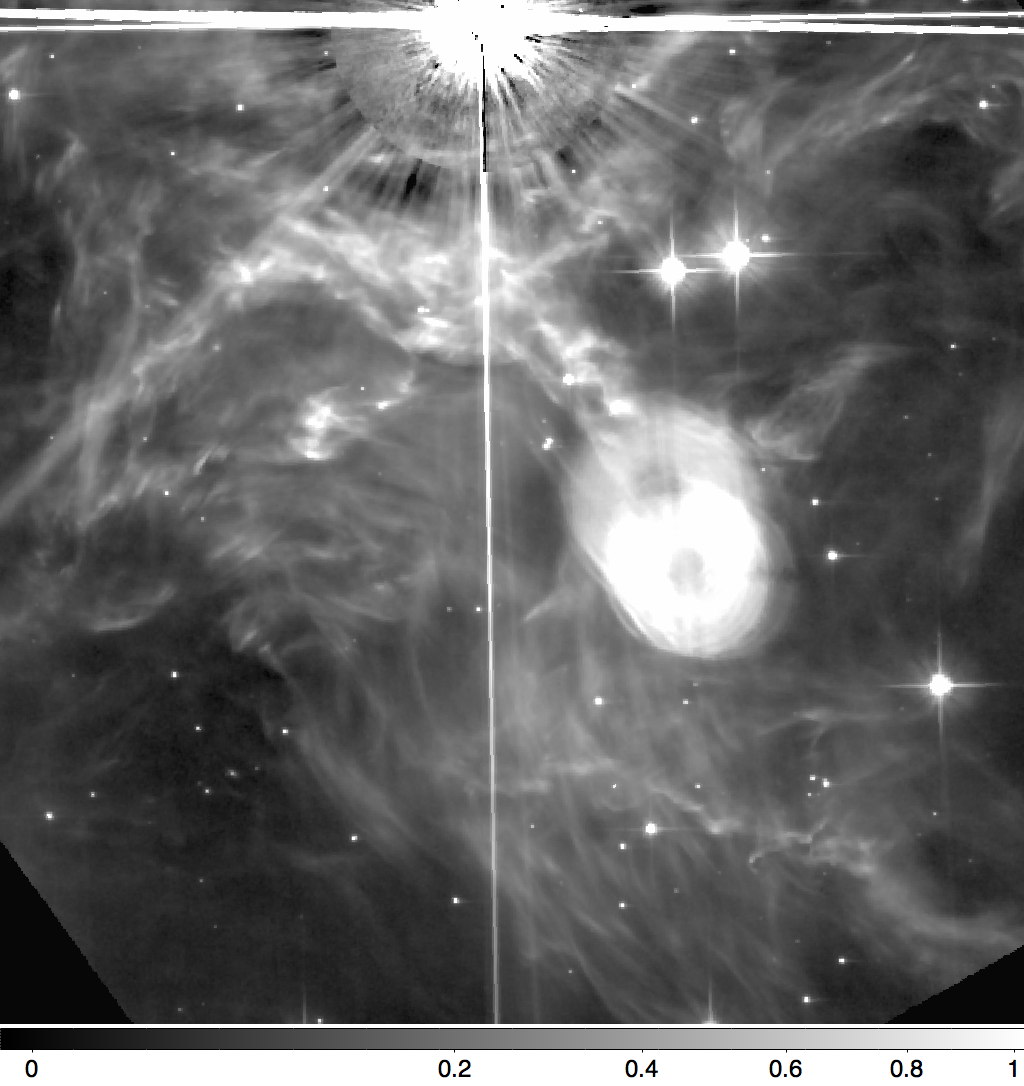} 
\includegraphics[width=6cm]{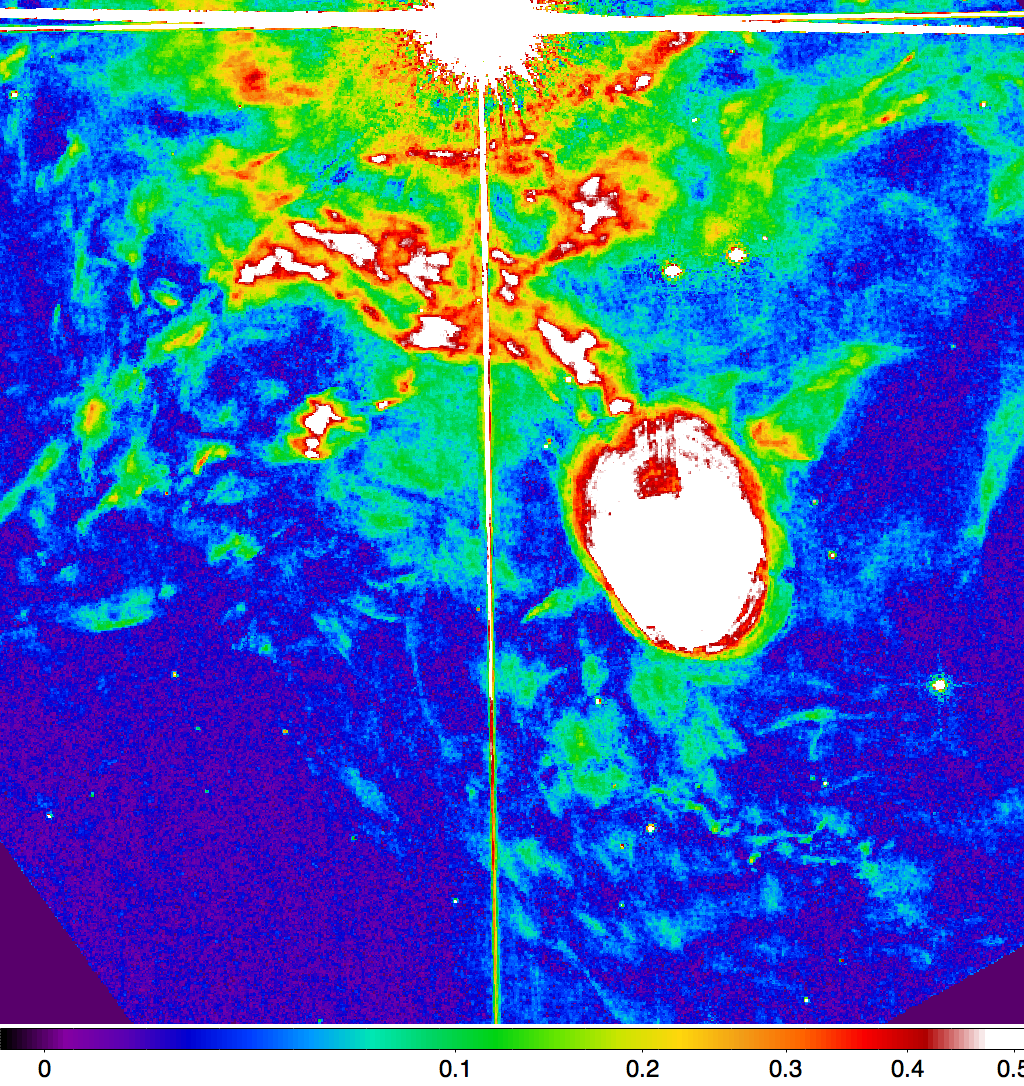}
\includegraphics[width=6cm]{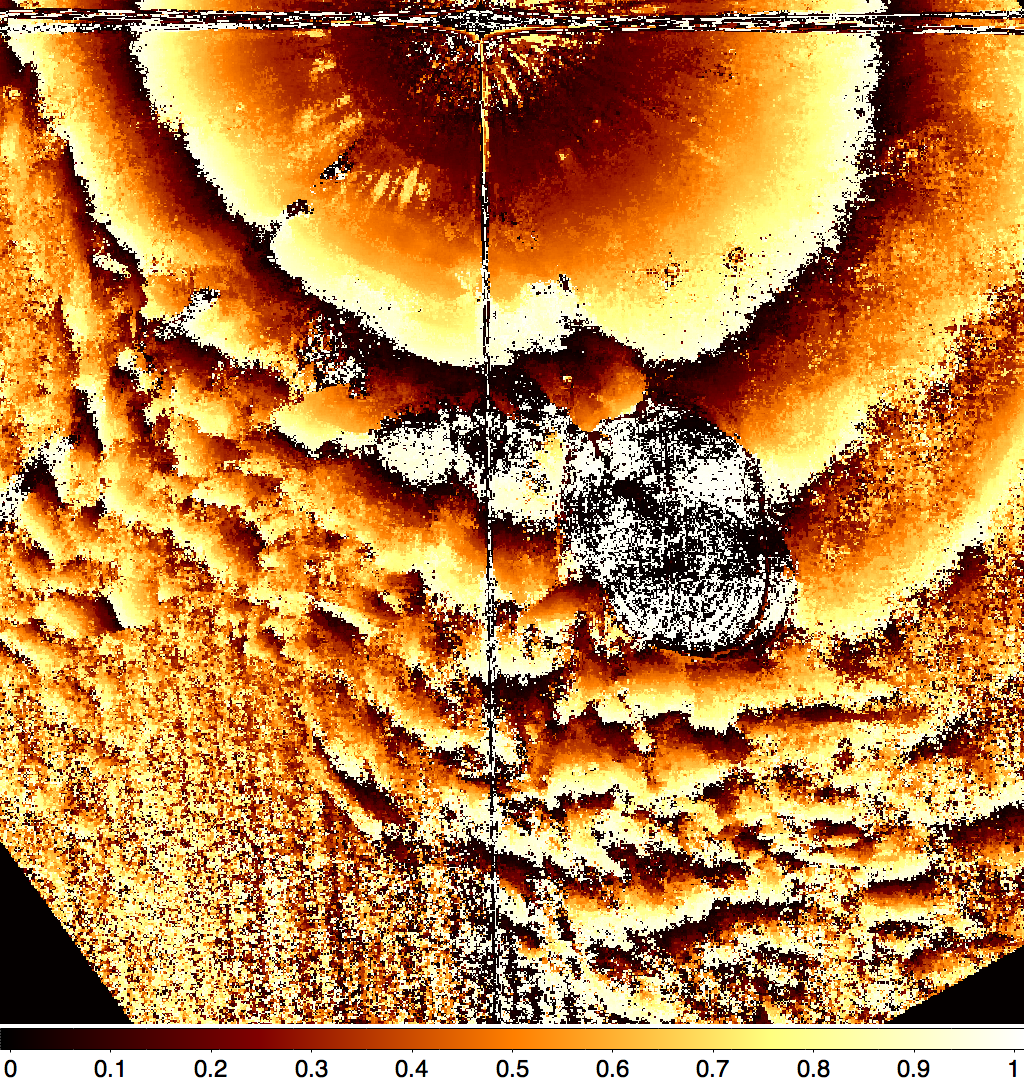}
\caption{Maps of the average flux $f_0$ (left panel, in e$^-$\ s$^{-1}$), amplitude $A$ (central panel, in e$^-$\ s$^{-1}$) and phase offset $\Delta \phi$ (right panel) of the photometric variations observed in the Southern quadrant of the nebula of RS~Pup. The FoV is $68.3'' \times 68.3''$, with North up and East to the left.\label{3paramFigure}}
\end{figure*}

As discussed by \citetads{2012A&A...541A..18K} and \citetads{1972A&A....16..252H}, the scattering layer must be geometrically thin as we observe contrasted light echoes over most of the nebula. But its non-zero thickness nevertheless causes the shape of the light curve to change over the nebula. A very thin layer produces a high contrast curve that is mostly identical to the light curve of the Cepheid, while a thicker dust layer will result in phase smearing and a smoother, lower amplitude curve with a shifted maximum flux phase. Such a shift of the maximum flux phase $\Delta \phi$ could affect our distance determination.
We therefore computed a second photometric fit using smoothed versions of  the light curve of RS Pup for those regions of the nebula where the first fit yielded a reduced $\chi^2 > 1$.
These light curves were computed using a boxcar moving average. We increased progressively the boxcar width from 0.1 to 1.0 in phase (with a 0.1 step), stopping for each pixel when the reduced $\chi^2$ became lower than 1. This process produced new triplets $(A, f_0, \Delta \phi)$ for each point over the nebula, as well as a value of the boxcar width. We note that this second fitting step effectively reduces the $\chi^2$ of the 3-parameter model fit, but its effect is small in terms of phase offset over most of the nebula.
The resulting maps of the average flux $f_0$, amplitude $A$ and phase offset $\Delta \phi$ are presented in Fig.~\ref{3paramFigure}.

\section{Polarization model \label{polarmodel}}

\begin{figure}[]
\centering
\includegraphics[width=\hsize]{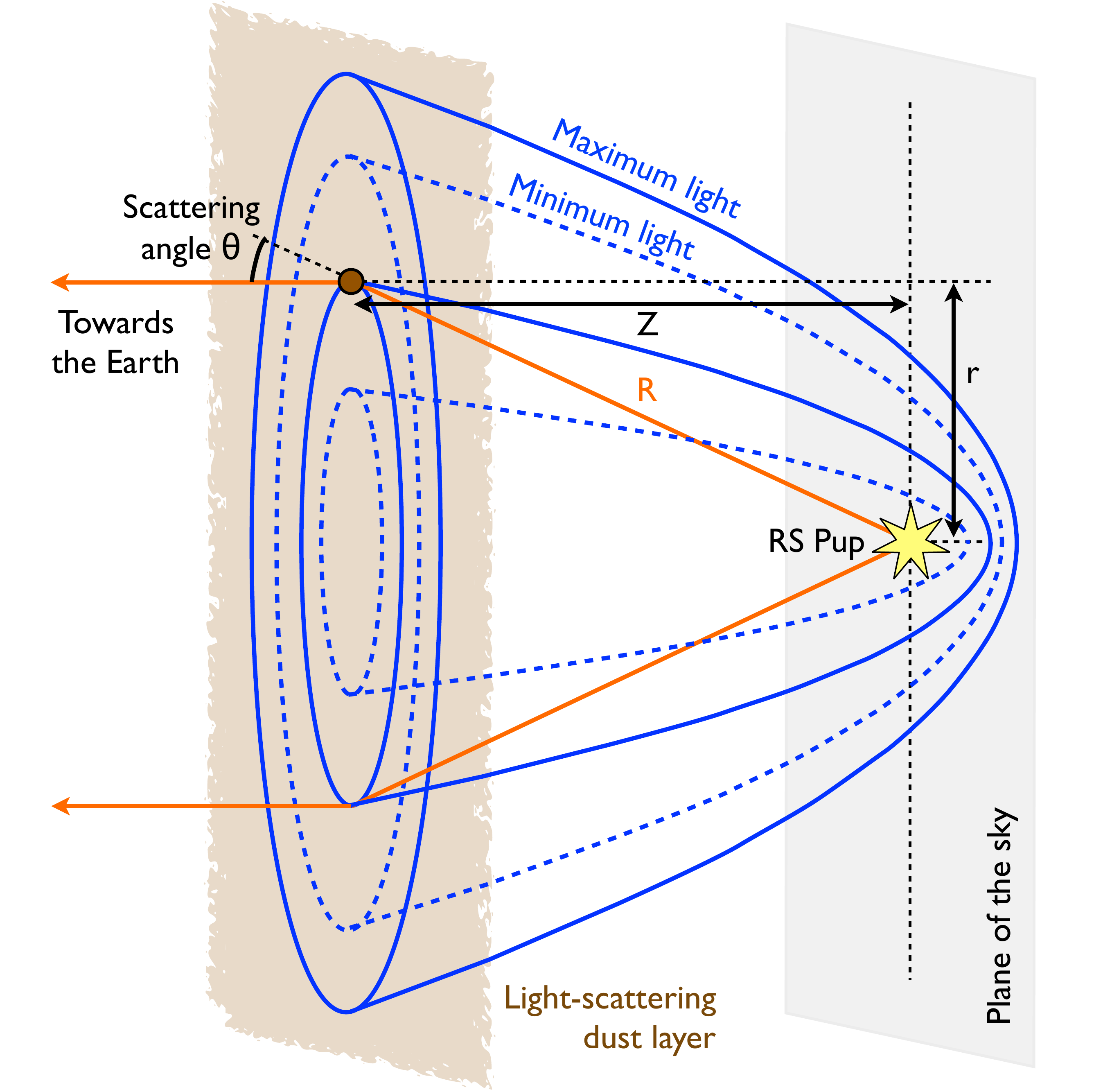}
\caption{Geometry of the light echoes in the dust surrounding RS~Pup.\label{Figure_Geometry}}
\end{figure}

The overall geometrical configuration of the scattering layer is shown in Fig.~\ref{Figure_Geometry}. The computation of the position of the dust layer with respect to the star is based on the determination of the scattering angle $\theta$ from the degree of linear polarization $p_L$. We here consider two models: (1) a physical model of the Milky Way dust developed by \citetads{2003ApJ...598.1017D}, scaled to the observed $p_\mathrm{max}$ value, and (2) a parametrized Rayleigh scattering model with two parameters $(p_\mathrm{max}, \theta_\mathrm{max})$. 

\subsection{Maximum degree of linear polarization $p_\mathrm{max}$ \label{pmaxvalue}}

\begin{figure}[]
\centering
\includegraphics[width=\hsize]{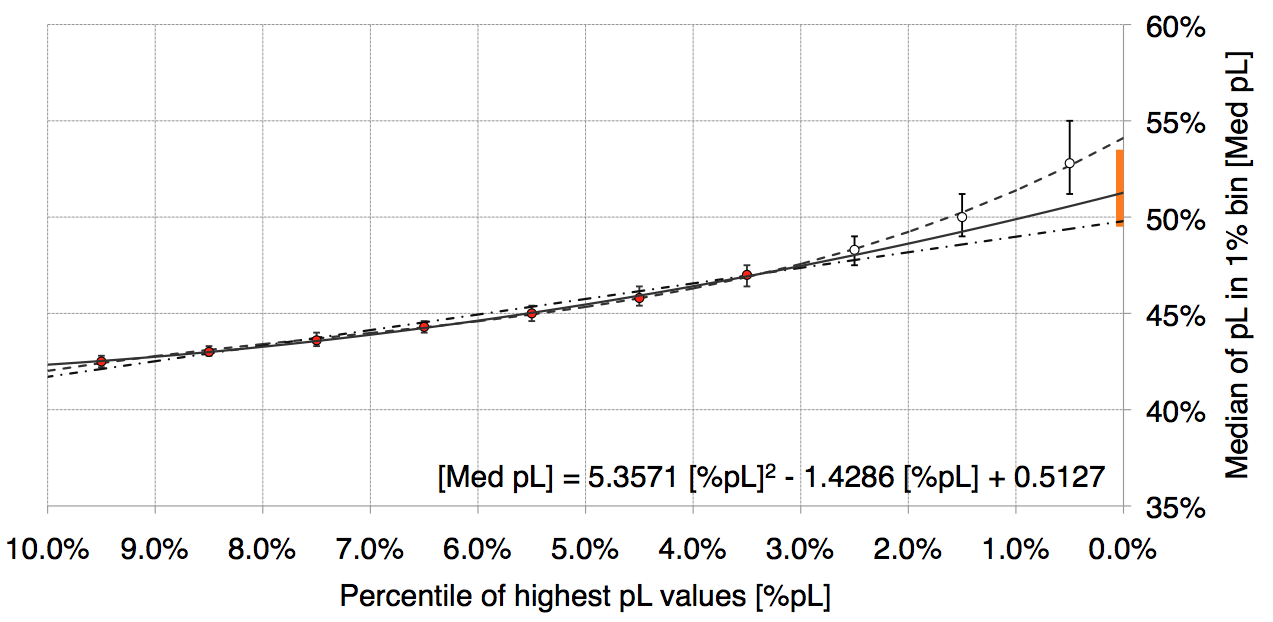}
\caption{Median $p_L$ values observed in the nebula of RS~Pup in 1\% bins, as a function of the percentile of highest values. The attached vertical segments show the full range of values of $p_L$ in each bin.  The orange segment on the vertical axis shows the adopted uncertainty range for $p_\mathrm{max} = \pmax\% \pm \pmaxerr\%$.
\label{MaxPolCurve}}
\end{figure}

For both polarization models (Rayleigh and Draine), the maximum degree of polarization $p_\mathrm{max}$ can be estimated from our ACS observations. 
But in practice, the determination of the maximum value of a statistical distribution is a relatively difficult task. We cannot simply read the maximum value of $p_L$ in the pixels of our map ($\approx 55\%$), as the measurements are affected by a statistical uncertainty. We would therefore obtain a too high value, biased by the statistical dispersion.
To estimate the true value of $p_\mathrm{max}$, we analyzed the statistical distribution of the values of $p_L$ in the nebula. We sorted the measured values of $p_L$ by increasing value, and binned them into percentile bins of width 1\%.
The sample we considered contains 21\,600 pixels with an SNR on $p_L$ larger than 10, i.e.~216\,pixels per percentile bin.
A third degree polynomial fit to the 0-10\% bin median values is shown in Fig.~\ref{MaxPolCurve} as a dashed curve. Its intercept at 100\% percentile gives a $p_\mathrm{max}=54\%$. However, the highest percentile bins (0-3\%) show a deviation compared to the general linear trend that is present over the 3-10\% percentiles.
The reason for this deviation is that the average statistical error bar of each $p_L$ measurement in our sample is $\sigma = 3.7\%$. This means that if we consider the 1\% highest $p_L$ values that exhibit a 53\% median visibility, we sample essentially the positive statistical fluctuations of $p_L$ values that are in reality (on the sky) around 50\%. In other words, the median values in the highest percentiles (0-3\% corresponding to $p_L=48$ to 53\%) show the statistical ``tail'' of the gaussian distribution of the highest $p_L$ values rather than the true distribution of the $p_L$ values on the sky.

To avoid a bias on the measured value of $p_\mathrm{max}$, we therefore adopt a quadratic polynomial fit and ignore the three highest $p_L$ bins (0 to 3\%) which are shown with open symbols in Fig.~\ref{MaxPolCurve}.
The intercept of the quadratic polynomial fit to the resulting measurements is $p_\mathrm{max} = \pmax\%$, as shown by the solid curve in Fig.~\ref{MaxPolCurve}. To estimate the uncertainty in this fitting procedure, we also adjusted a linear model, which gives an intercept of 50\%.
Considering the range of results between the linear, quadratic and third degree polynomial fits, we adopt $p_\mathrm{max} = \pmax\% \pm \pmaxerr\%$ for our two polarization models (Milky Way dust and Rayleigh) that is shown as an orange segment in Fig.~\ref{MaxPolCurve}.

\begin{figure}[]
\centering
\includegraphics[width=\hsize]{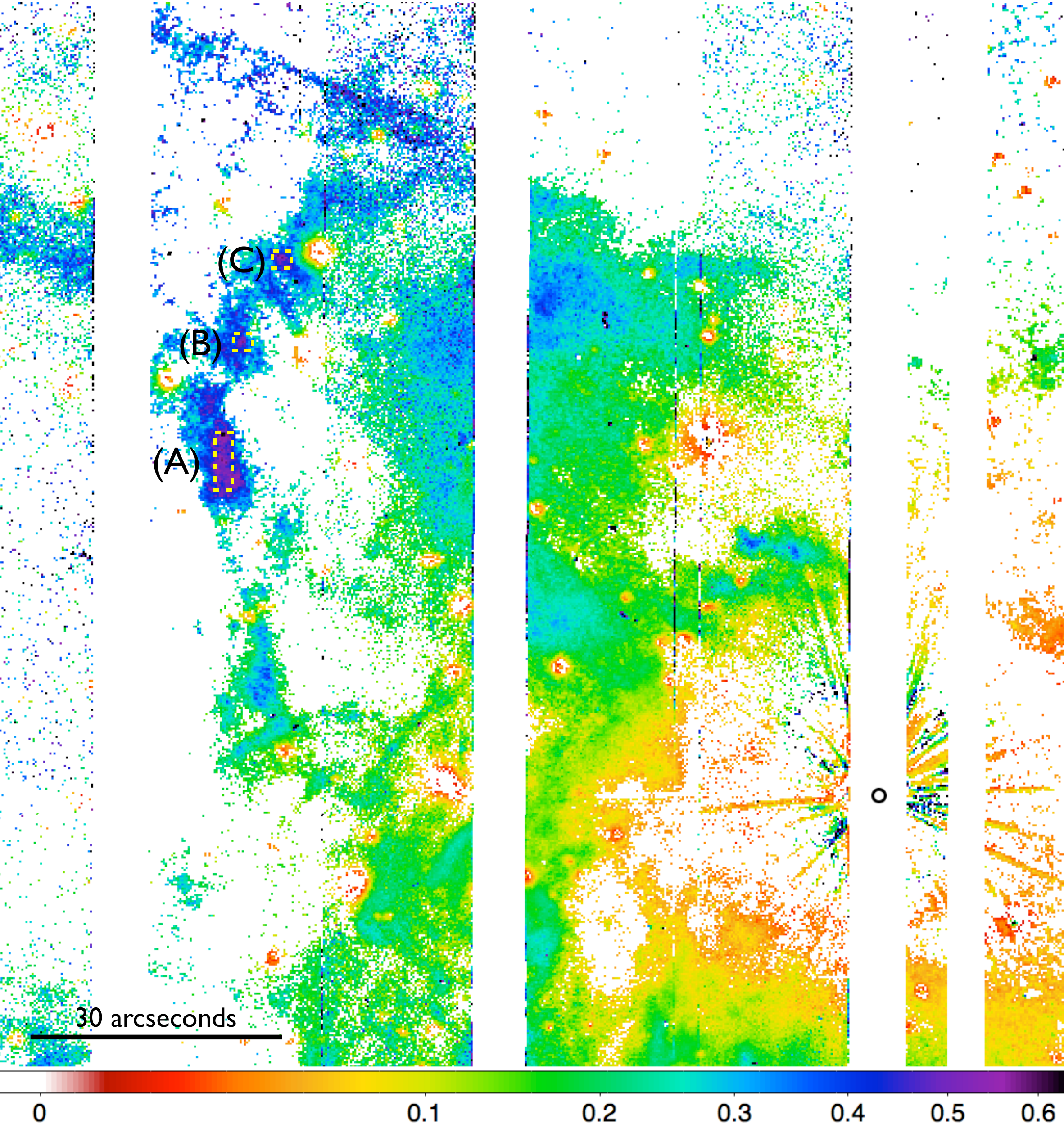}
\caption{Map of the degree of linear polarization $p_L$ in the nebula of RS~Pup from VLT/FORS observations  \citepads{2012A&A...541A..18K}.
\label{pMax-FORS}}
\end{figure}

We can compare this value with the observations by \citetads{2012A&A...541A..18K} using the VLT/FORS instrument. A map of the degree of linear polarization measured in the nebula of RS~Pup by these authors in the $V$ band is presented in Fig.~\ref{pMax-FORS}. The regions of highest polarization labeled (A), (B) and (C) are located at approximately $1.5\arcmin$ from the Cepheid. The median polarization and standard deviation of the $p_L$ values measured over these windows is $(A)=50.0\% \pm 3.6\%$, $(B)=50.2\% \pm 4.7\%$ and $(C)=49.0\% \pm 4.7\%$. These maximum values are in good agreement with the $p_\mathrm{max}$ value we measure in the HST/ACS images in the same wavelength range.
Another comparison is the maximum value of $p_L$ measured by \citetads{2008AJ....135..605S} in the circumstellar nebula of V838~Mon. These authors observed $p_\mathrm{max} \approx 50\%$ (see also Sect.~\ref{drainemodel} and \ref{literaturepola}), again in good agreement with our value. The case of the nebula of V838~Mon is particularly interesting as the dust is distributed relatively homogeneously around the star. This means that the maximum polarization $p_\mathrm{max}$ is actually reached in the nebula, as all scattering angles are present in the echo. 

The value of $p_\mathrm{max}$ that we derive is formally a lower limit, as all scattering angles may not be present in the ACS FoV. However, we make here the hypothesis that a fraction of the dust we observe is present down to the plane of the sky and beyond, i.e. that it is not all confined between us and the Cepheid. Considering the overall shape of the nebula and of its distant extensions (Fig.~\ref{pMax-FORS}) this assumption appears reasonable. We note that we have not observed $p_L$ values both in our VLT/FORS and HST/ACS larger than $p_\mathrm{max} = \pmax\%$ in a statistically significant way. We therefore consider our value of $p_\mathrm{max}$ in the following as a measurement and not as a lower limit.

\subsection{Milky Way dust polarization model \label{drainemodel}}

The Milky Way (MW) dust model presented in Fig.~5\footnote{Data available from \url{https://www.astro.princeton.edu/~draine/dust/scat.html}} of \citetads{2003ApJ...598.1017D} (for $R_V=3.1$) gives $p_\mathrm{max}=25.5\%$ for $\theta_\mathrm{max}=90^\circ$ at $616.5$\,nm. This wavelength is the closest to the ACS WFC F606W filter, as the range covered by this filter is $\approx 470-710$\,nm, with $\lambda_C = 590$\,nm and $\Delta \lambda = 230$\,nm.

The difference between the $p_\mathrm{max}$ value of the MW dust model ($24-27\%$) with what we observe (\pmax\%) points at a difference in the dust grain properties. In the theoretical Rayleigh scattering case, for which the typical size of the particles $R$ is small compared to the wavelength $\lambda$, we expect that $p_\mathrm{max} \rightarrow 100\%$ and $\theta_\mathrm{max} \rightarrow 90^\circ$, with $p_\mathrm{max}$ decreasing with increasing $R/\lambda$. Although the variation may not be strictly monotonic for large grains \citepads{2009ApJ...696.2126S}, the high observed $p_\mathrm{max}$ indicates that the dust grains surrounding RS~Pup and V838 Mon are probably smaller in size than the average MW mixture considered by \citetads{2003ApJ...598.1017D}.

\begin{figure}[]
\centering
\includegraphics[width=\hsize]{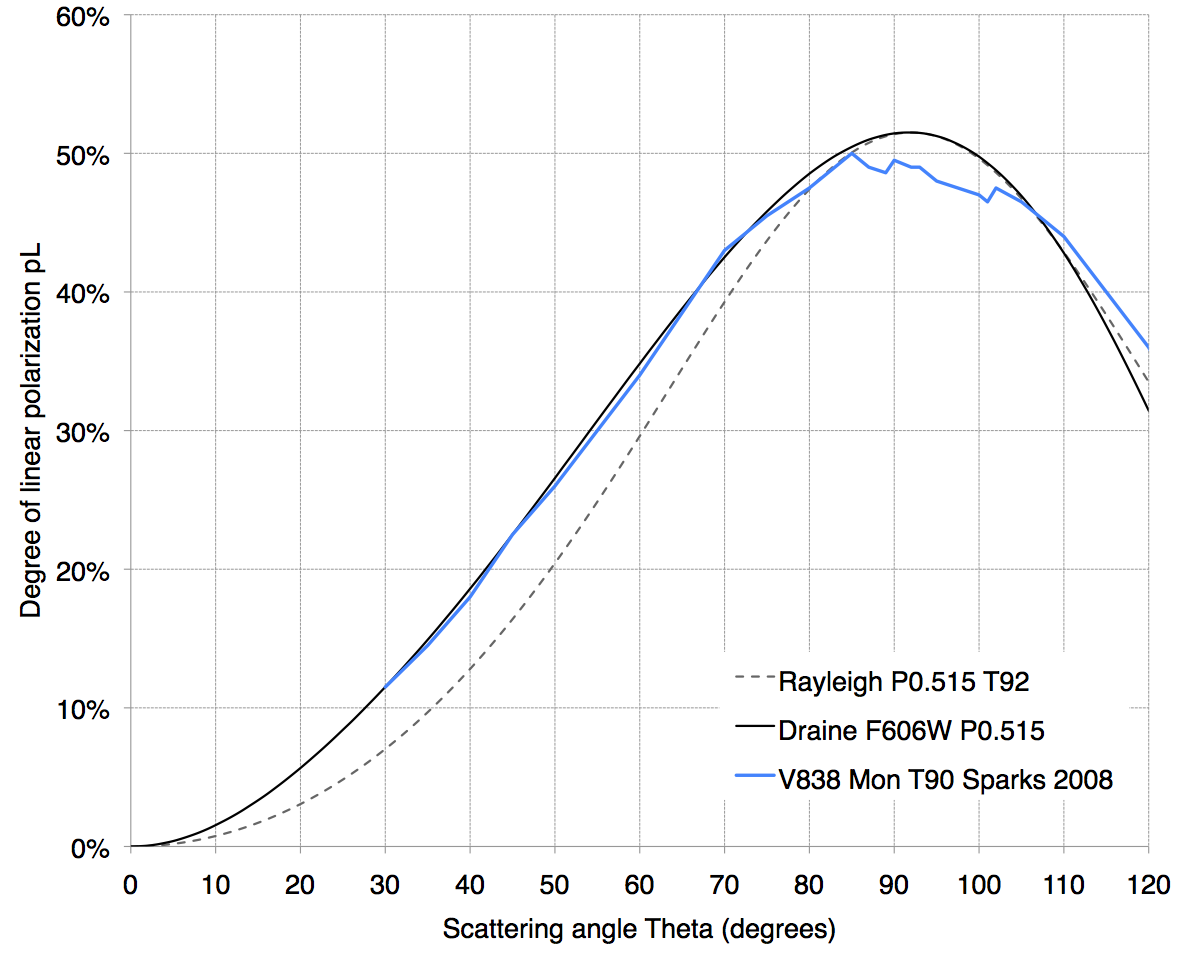}
\caption{Milky Way dust ($p_\mathrm{max} = \pmax\%$, solid black curve) and Rayleigh ($p_\mathrm{max} = \pmax\%$, $\theta_\mathrm{max} = \thetamax^\circ$, dashed curve) models of the linear polarization degree $p_L(\theta)$. The observed polarization function measured by \citetads{2008AJ....135..605S} for the dust surrounding V838~Mon is shown as a blue curve. \label{polarizationModel}}
\end{figure}

We derive a suitable polarization model for our observations by scaling the MW dust models to match the $p_\mathrm{max} = \pmax\%$ value that we measure on the nebula of RS~Pup. We took into account the bandpass of the F606W filter by computing the average of the relevant polarization models from \citetads{2003ApJ...598.1017D}, weighted as a function of wavelength by the photometric throughput of the ACS WFC+F606W as listed in the ACS Handbook \citepads{2012acsi.book.....U}. We thus obtained the $p_L(\theta)$ model shown in Fig.~\ref{polarizationModel}.
It is interesting to note that the model we derive using our independently determined value of $p_\mathrm{max}$ (Fig.~\ref{polarizationModel}) provides an excellent match to the observed polarization function of V838~Mon derived by \citetads{2008AJ....135..605S}.

\subsection{Rayleigh polarization model \label{rayleighmodel}}
The dependence of the linear polarization degree $p_L$ on the scattering angle $\theta$ may be assumed to be given by the classical Rayleigh polarization phase function \citepads{1979ApJ...229..954W}. For theoretical Rayleigh scattering, $p_\mathrm{max}=100\%$ is reached in the plane of the sky, i.e.~the imaginary plane perpendicular to the line of sight containing RS~Pup, for which $\theta_\mathrm{max} = 90^\circ$ (Fig.~\ref{Figure_Geometry}). However, the maximum values of $p_L$ observed in astrophysical context are lower than 100\%, and the angle of maximum polarization may be different from $90^\circ$.
We introduce two parameters to account for possible deviations of the actual dust scattering properties from the Rayleigh polarization law, the maximum degree of linear polarization $p_\mathrm{max}$ and the angle of the maximum polarization $\theta_\mathrm{max}$. This is the approach selected by \citetads{2008AJ....135..605S} for their study of V838~Mon's light echo. The expression of $p_L$ can be written as:
\begin{equation}\label{pl_eq}
p_L = p_{\rm max}\,\left( \frac{1-\cos^2 \tilde{\theta} }{1+\cos^2 \tilde{\theta} } \right)
\end{equation}
where $p_{\rm max}$ is the maximum degree of linear polarization, and $\tilde{\theta}$ is the scaled scattering angle defined as 
\begin{equation}\label{theta_eq}
\tilde{\theta} = \frac{\theta}{\theta_\mathrm{max}} \times 90^\circ
\end{equation}
The $p_{\rm L}(\theta, p_\mathrm{max}, \theta_\mathrm{max})$ function reaches its maximum value $p_\mathrm{max}$ for $\theta = \theta_\mathrm{max}$ (Fig.~\ref{polarizationModel}). For our Rayleigh polarization model, we adopt the measured value of $p_\mathrm{max} = \pmax\% \pm \pmaxerr\%$ (Sect.~\ref{pmaxvalue}).

The second parameter $\theta_\mathrm{max}$ of our polarization model cannot be measured directly from the observations.  This
parameter, estimated by \citetads{2008AJ....135..605S} to be in the range $90^\circ \pm 5^\circ$, is the main source of systematic uncertainty in their distance estimate of V838~Mon.
 From our broadband MW dust model presented in Sect.~\ref{drainemodel}, we derive $\theta_\mathrm{max}(\mathrm{MW, F606W}) = \thetamax^\circ$. Over the full visible range ($0.4-0.8\,\mu$m), the value of $\theta_\mathrm{max}$ for the models by \citetads{2003ApJ...598.1017D} varies between $90^\circ$ and $95^\circ$, with a standard deviation of $\approx 2^\circ$.
We thus adopt an uncertainty range of $\theta_\mathrm{max} = \thetamax^\circ \pm 2^\circ$ for our Rayleigh polarization model of RS~Pup's nebula.

\subsection{Polarization models from the literature \label{literaturepola}}

As mentioned above, the polarization model is the leading systematic uncertainty in this investigation. It is thus useful to compare the model adopted here with others from the literature.

For their study of the Homunculus nebula surrounding $\eta$\,Car, \citetads{1999AJ....118.1320S} assumed a $p_L(\theta)$ function peaking around $\theta_\mathrm{max} \approx 97^\circ$ at $p_\mathrm{max} \approx 34\%$ in the $V$ band (WFPC2+F555W). This assumed value of $\theta_\mathrm{max}$ is $2.5 \sigma$ away from our scaled MW dust model, but the actual function depends on the composition of the dust that is likely different in the Homunculus and the average MW interstellar medium.

The massive nebula surrounding the red supergiant VY\,CMa was observed in polarimetric imaging by \citetads{2007AJ....133.2730J} using \emph{HST}/WFPC2 and the F550M and F658N (H$\alpha$) filters. The measured $p_L$ values reach 75\%, with an assumed $p_L(\theta)$ function peaking around $\theta_\mathrm{max}=100^\circ$.

The observations of the circumstellar disk of AU~Mic by \citetads{2007ApJ...654..595G}  using \emph{HST}/ACS in the F606W filter were well reproduced by a Rayleigh polarization function peaking at $\theta_\mathrm{max}=90^\circ$. These authors fitted simultaneously the \citetads[][hereafter H-G]{1940AnAp....3..117H, 1941ApJ....93...70H} phase function and the linear polarization function $p_L(\theta)$ to the data. They obtained $p_\mathrm{max} = 53 \pm 3\%$ for their single-component H-G model with a strong forward scattering, in good agreement with the value we derive for RS~Pup.
\citetads{1999AJ....117.1408S} observed the proto-planetary nebula Roberts 22 using \emph{HST}/WFPC2 polarimetry in the F606W filter. They measure comparable maximum $p_L$ values ($40-50\%$) as what we find in the nebula of RS~Pup, and they reproduce their observations with a distribution of low albedo grains and a Rayleigh scattering phase function ($\theta_\mathrm{max} = 90^\circ$).

The fact that we observe a very similar value of $p_\mathrm{max}$ in RS~Pup's nebula as \citetads{2008AJ....135..605S} in V838~Mon gives us confidence that the dust surrounding these two stars share similar physical properties. As shown by \citetads{2012A&A...548A..23T} and \citetads{2012A&A...541A..18K}, these dust clouds were not created by mass loss from the two stars, but are the remnants of interstellar dust clouds.
The dust envelopes of $\eta$\,Car and VY\,CMa have a different origin, as they were recently formed by mass loss from the central stars.

The experiments by \citetads{2007A&A...470..377V} at a wavelength of 633\,nm show very high degrees of linear polarization (up to $p_L \approx 50\%$ to 100\%) on light scattered by fluffy dust grains. Such grains are expected to be good analogs of the interstellar dust. The high $p_L$ values are interpreted as the signature of the scattering from the small-size grains in the aggregates, while the phase function shape is determined by the size distribution of the aggregates. The measured maximum polarization angle $\theta_\mathrm{max}$ values in the different samples studied by these authors are consistently very close to $90^\circ$. The high $p_L$ values observed in V838~Mon and RS~Pup's nebulae therefore point at the presence of such aggregates of very small grains and lend support to our adopted range of $\thetamax^\circ \pm 2^\circ$ for $\theta_\mathrm{max}$.
Photochemical processing of the dust grains as they are heated by the central star's radiation may also lead to the destruction of the larger, porous and fluffy grains to form a population of very small grains with scattering properties closer to the Rayleigh hypothesis. This process was probably more efficient in the past than it is now, as the main sequence progenitor of RS~Pup was a hot B-type star producing strong ultraviolet radiation.

\section{Distance determination from the light echoes \label{distanceFit}}

\subsection{Selection of suitable nebular features \label{selection}}

\begin{figure}[]
\centering
\includegraphics[height=\hsize]{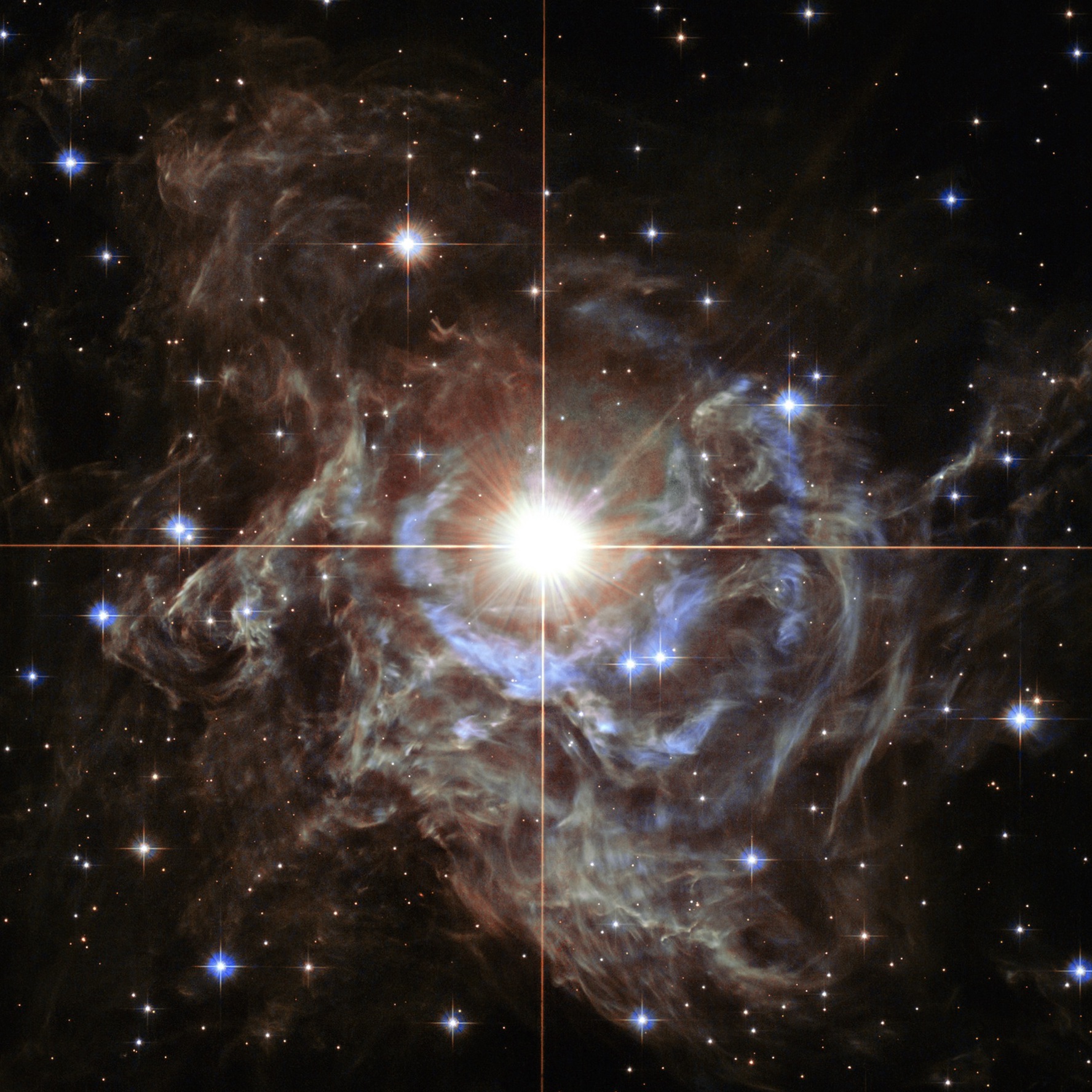}
\caption{Color composite view of the circumstellar nebula of RS\,Pup assembled from the ACS WFC images in the F435W and F606W filters obtained on 2010 March 26 (color rendition: NASA, ESA, Z.~Levay, and the Hubble Heritage Team STScI/AURA-Hubble/Europe Collaboration). The maximum light echoes are visible as blue rings due to the hotter temperature of the central star around its maximum light. North is up and East to the left, and the FoV is $2.9 \times 2.9$\,arcminutes.\label{rspupcolor}}
\end{figure}

The overall morphology of the RS Pup nebula is relatively complex (Fig.~\ref{rspupcolor}), with a number of knots and filaments spread over the apparent extent of the dust. To obtain a higher signal-to-noise ratio (SNR) on the $p_L$ map, we averaged the seven observing epochs and binned the resulting map by $2 \times 2$\,pixels. A temporal variability of $p_L$ is detectable between our seven observing epochs. This is caused by the propagation of the parabolic light echo surfaces (Fig.~\ref{Figure_Geometry}) into the 3-dimensional circumstellar material, which results in different scattering angles as a function of time (hence different values of $p_L$). This variability is observed mostly in the low polarization regions (close to the central star), which do not constrain properly the altitude of the scattering material above the plane of the sky. For this reason, they do not affect our distance fit, and we considered the average map of $p_L$ for our distance analysis.

\begin{figure}[]
\centering
\includegraphics[width=\hsize]{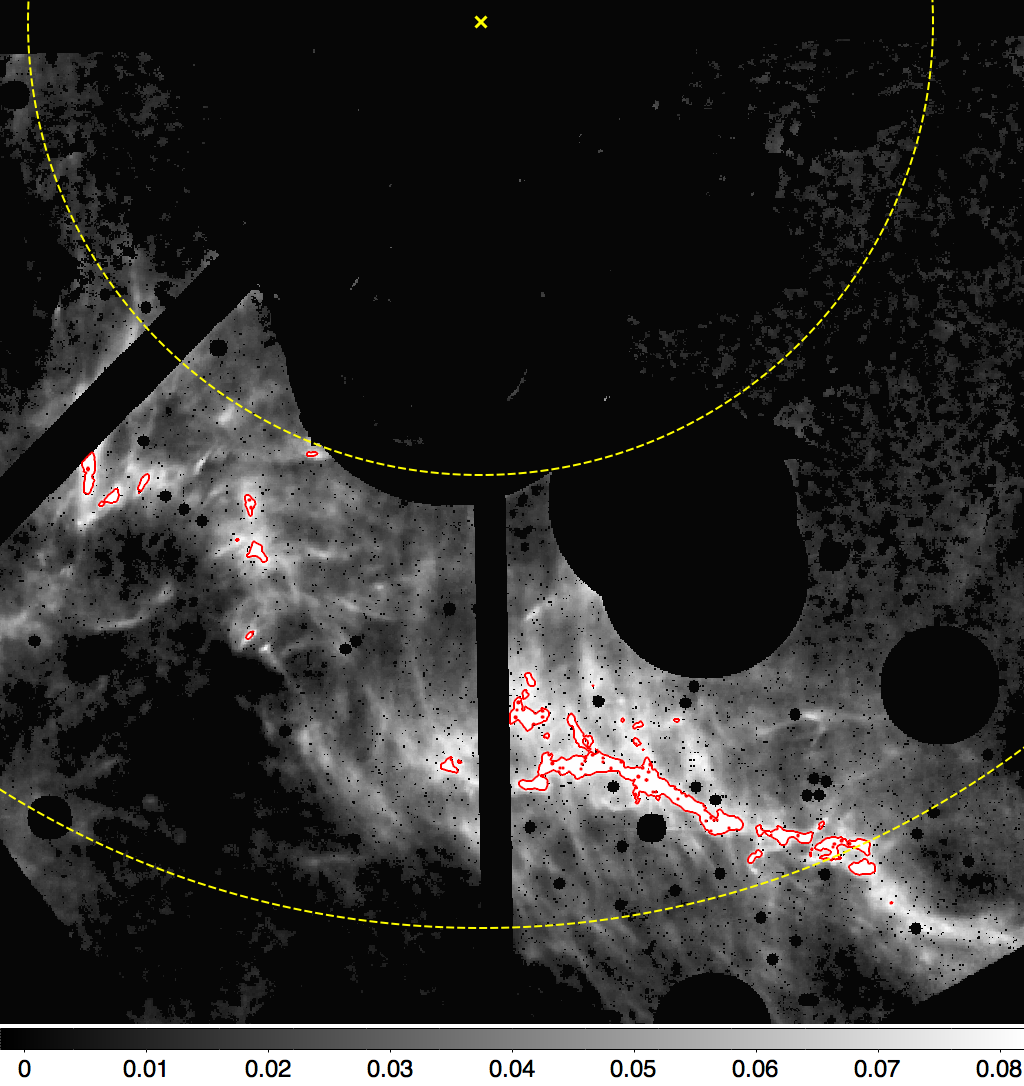} 
\caption{Map of the polarized flux $I \times p_L$ in the nebula of RS~Pup (in $e^-\ s^{-1}$). The red contours enclose the regions that were selected for the distance fit. The artefacts and background stars have been masked, and the yellow cross marks the position of RS~Pup. The dashed yellow circles mark $30''$ and $60''$ angular separations from the Cepheid, and the FoV is $68.3'' \times 68.3''$ with North up and East to the left. \label{IxPLmap}}
\end{figure}

For the selection of the suitable areas of the RS~Pup images for the distance fit, we first determined the FoV limits (the pointing of the telescope slightly changes between epochs), and excluded all the regions that are affected by artefacts (pupil ghosts, diffraction spikes,...), as well as the background stars and their associated halos. We also masked the pixels for which the 3-parameter fit (Sect.~\ref{3paramFit}) did not converge properly. This resulted in a fraction of selected pixels corresponding to 48\% of the full ACS frame. After smoothing the $p_L$ map using a $7 \times 7$\,pixel box, we selected the regions that have a sufficiently high polarimetric SNR $p_L/\sigma(p_L) \geqslant 10$. This resulted in the selection of 25\% of the original pixels. We then selected the highest polarized flux pixels, i.e.~the pixels for which $I \times p_L \geqslant 0.08\,e^-$ (Fig.~\ref{IxPLmap}). This selection ensures both that the selected regions are sufficiently high above the noise background (high $I$), and that the scattering occurs essentially in a single dust layer located relatively close to the plane of the sky (high $p_L$). This selection process resulted in the identification of 4000 suitable pixels ($\approx 0.4\%$ of the full ACS frame with $2\times2$ binning).

While the presence of contrasted light echoes over most of the nebula indicates that the scattering layer is generally thin, the observed behavior of the echoes varies depending on the location. For instance, the relative amplitude of the photometric variation  in the nebula is lower than the Cepheid's own variation. When this amplitude is much lower than the Cepheid's one, the dust may be distributed in a thick layer or even in several superposed layers. In this case, the derivation of a single phase lag is difficult and may lead to biases. Therefore, we tried selecting pixels in the nebula that show a high relative photometric variation amplitude. However this procedure selected essentially only pixels with high values of $I \times p_L$ value. We therefore preferred to keep the selection process simple in order to avoid biases, and we did not apply this additional selection criterion.

\subsection{Light echoes model}

When fitting for the distance, we employ the measured $p_L$ map to infer the scattering angle $\theta(p_L)$ from both the Milky Way and Rayleigh polarization models defined in Sect.~\ref{polarmodel}. We assume in the following that the scattering angle is lower than or equal to $\theta_\mathrm{max}$, i.e.~that the scattering occurs in the forward direction. At visible wavelengths, forward scattering is far more efficient than backward scattering, which means that the dust we observe around RS~Pup is essentially located between us and the Cepheid, rather than behind the star. We also consider that the physical size of the nebula ($\approx 1$\,pc) is negligible with respect to the distance to RS~Pup ($\approx 2$\,kpc).
The altitude $Z$ of the scattering dust above the plane of the sky (as shown in Fig.~\ref{Figure_Geometry}) is derived through:
\begin{equation}
Z_{[\mathrm{pix}]}(p_\mathrm{max}, \theta_\mathrm{max}) = \tan\left[ 90^\circ - \theta(p_L, p_\mathrm{max}, \theta_\mathrm{max}) \right]\ r_{[\mathrm{pix}]}
\end{equation}
where $\theta(p_L, p_\mathrm{max}, \theta_\mathrm{max})$ is the selected polarization model, and $r_{[\mathrm{pix}]}$ the apparent angular separation between the dust feature and the central star. For the MW dust model (Sect.~\ref{drainemodel}), we consider $p_\mathrm{max} = \pmax\% \pm \pmaxerr\%$ and $\theta_\mathrm{max}$ is not a free parameter. For the Rayleigh model (Sect.~\ref{rayleighmodel}), we additionally assume $\theta_\mathrm{max} =\thetamax^\circ \pm 2^\circ$. The resulting altitude $Z$ is in the same physical unit as the projected separation $r$ expressed here in pixels for convenience. 

The linear length $C_{[\mathrm{m}]}$ of the variation cycle of RS~Pup propagating in space with the velocity $c$ is given by $C_{[\mathrm{m}]} = P_{[\mathrm{s}]}\,c_{[\mathrm{m\,s}^{-1}]}$ where $P = 41.5117$\,days is the photometric variation period.
The linear size $\ell_{[\mathrm{m}]}$ corresponding to one pixel at RS~Pup's distance $d$ is given by
\begin{equation}
\ell_{[\mathrm{m}]}(d) = \epsilon_{[\mathrm{''}]}\ d_{[\mathrm{pc}]}\ \mathrm{AU}_{[\mathrm{m}]}
\end{equation}
where $\epsilon = 0.0993\,\mathrm{arcsec}\ \mathrm{pix}^{-1}$ is the plate scale of the $2 \times 2$\,pixel binned ACS images, $d_{[\mathrm{pc}]}$ is the distance to RS~Pup in parsecs, and $\mathrm{AU}_{[\mathrm{m}]}$ represents the length of one astronomical unit expressed in meters.
The angular scale $C_{[\mathrm{pix}]}$ (expressed in pixels) of one variation cycle at RS~Pup's distance $d$, is then given by the expression:
\begin{equation}
C_{[\mathrm{pix}]}(d) = \frac{C_{[\mathrm{m}]}}{\ell_{[\mathrm{m}]}} = \frac{C_{[\mathrm{m}]}}{\epsilon_{[\mathrm{''}]}\ d_{[\mathrm{pc}]}\ \mathrm{AU}_{[\mathrm{m}]}} 
\end{equation}

The linear radius $R(x,y)$ between RS~Pup located at position $(x_\star, y_\star)$ and the dust at the apparent position $(x,y)$ in the nebula is then computed from
\begin{equation}
R_{[\mathrm{pix}]}(x,y, p_\mathrm{max}, \theta_\mathrm{max}) = \sqrt{ \left( x-x_\star \right)^2 + \left(y-y_\star\right)^2 + Z(x,y)_{[\mathrm{pix}]}^2}.
\end{equation}
As discussed by \citetads{1939AnAp....2..271C} \citepads[see also][]{2003AJ....126.1939S, 2009A&A...495..371B}, the model phase lag $\Delta \phi(x,y)$ of the photometric variation of the dust is given by:
\begin{equation}
\Delta \phi_\mathrm{model}(x,y,p_\mathrm{max}, \theta_\mathrm{max},d) = \frac{R_{[\mathrm{pix}]}(x,y) - Z_{[\mathrm{pix}]}(x,y)}{C_{[\mathrm{pix}]}(d)}
\end{equation}

\subsection{Distance computation}

To derive the distance $d$ of RS~Pup, we minimize the following expression as a function of $d$:
\begin{equation}
\chi^2(d, p_\mathrm{max}, \theta_\mathrm{max}) = \sum_{(x,y) \in S}{\left[ \frac{\Delta \Phi}{\sigma(\Delta \phi)} \right]^2}
\end{equation}
where $S$ is the selected pixel sample as described in Sect.~\ref{selection} and
\begin{equation}
\Delta \Phi = \min \left( \{ \gamma \}; \{ \gamma+1 \}; \{ \gamma-1 \} \right).
\end{equation}
In this expression, the phase difference $\gamma$ is defined as $\gamma = \left| \Delta \phi_\mathrm{model} - \Delta \phi \right|$ and $\{...\}$ represents the fractional part. This procedure is necessary as the phase values derived from the ACS images are wrapped, i.e.~$\Delta \phi \in [0,1]$. Due to the complexity of the nebula and the fact that the dust distribution is not continuous, it is not possible to directly unwrap the measured phase lags from our data to retrieve the integral number $N$ of phase lag cycles \citepads[this was already the case for the study by][]{2012A&A...541A..18K}. We thus consider $\Delta \Phi$ as the distance between the model and the observations.
The uncertainties associated to the parameters of the polarization model $(p_\mathrm{max}, \theta_\mathrm{max})$ are systematic and we derive their contributions separately.
 
Before the computation of the $\chi^2$ minimization, we smoothed the polarization map using a $11 \times 11$\, pixel moving average, and the phase lag map using a $3\times3$\,pixel moving average to improve the SNR. Following \citetads{2012acsi.book.....U}, we adopt a minimum uncertainty on the degree of linear polarization $p_L$ of $\pm 1\%$, and we set a minimum uncertainty on the phase lag of $\pm 0.01$. In the highest photometric SNR regions of the nebula, the statistical uncertainties that we derive on these two quantities are very small, and fixing these minimum uncertainties ensure that we do not underestimate them.
 
As discussed in Sect.~\ref{literaturedistances}, the published distance estimates for RS~Pup range from $\approx 1800$\,pc \citepads{2011A&A...534A..94S} to $\approx 2100$\,pc \citepads{2003LNP...635...21F}. We therefore considered for the $\chi^2$ minimization a range of distances of 1500 to 2400\,pc.
 
\begin{figure*}[]
\centering
\includegraphics[width=9cm]{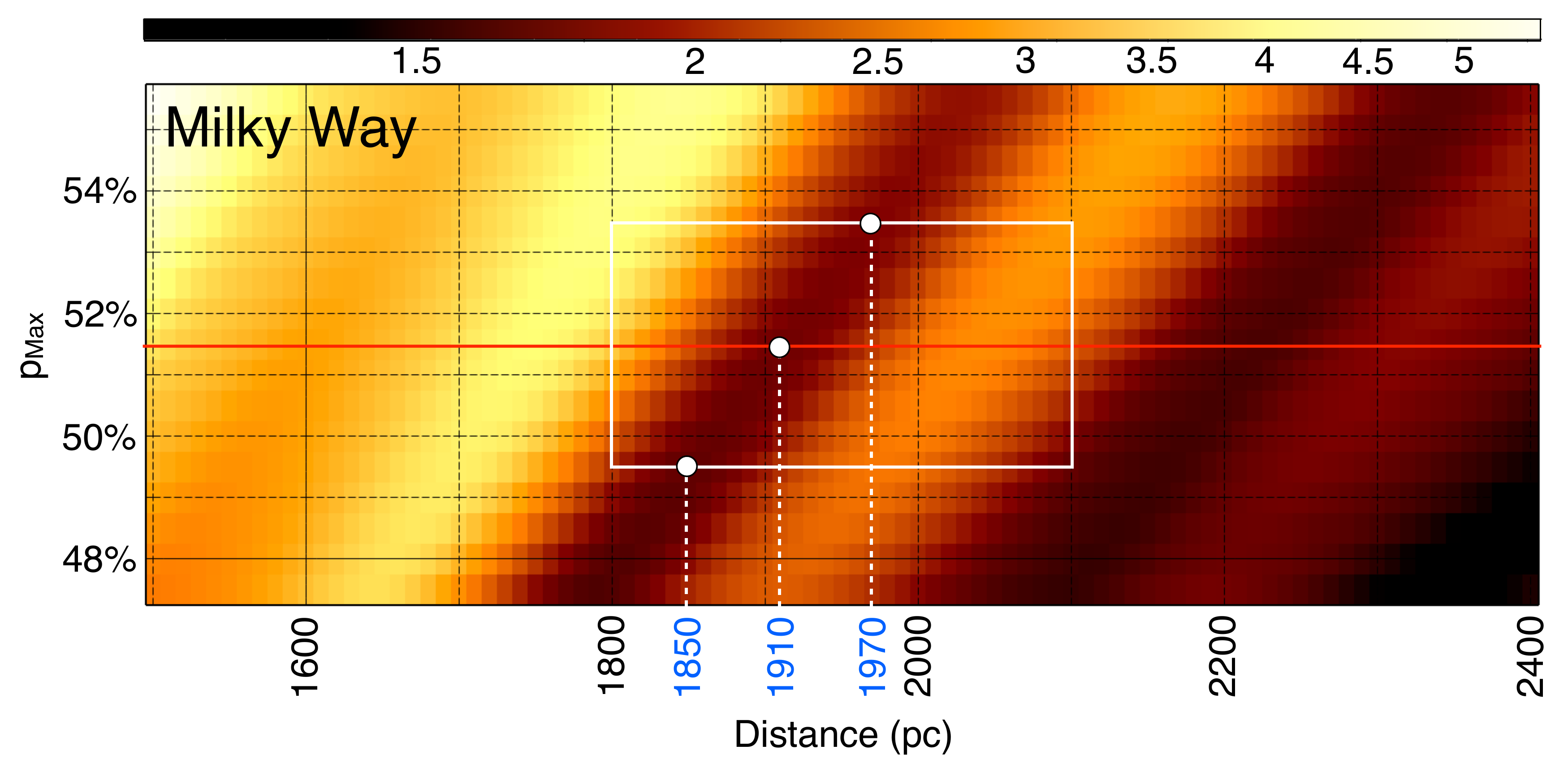} 
\includegraphics[width=9cm]{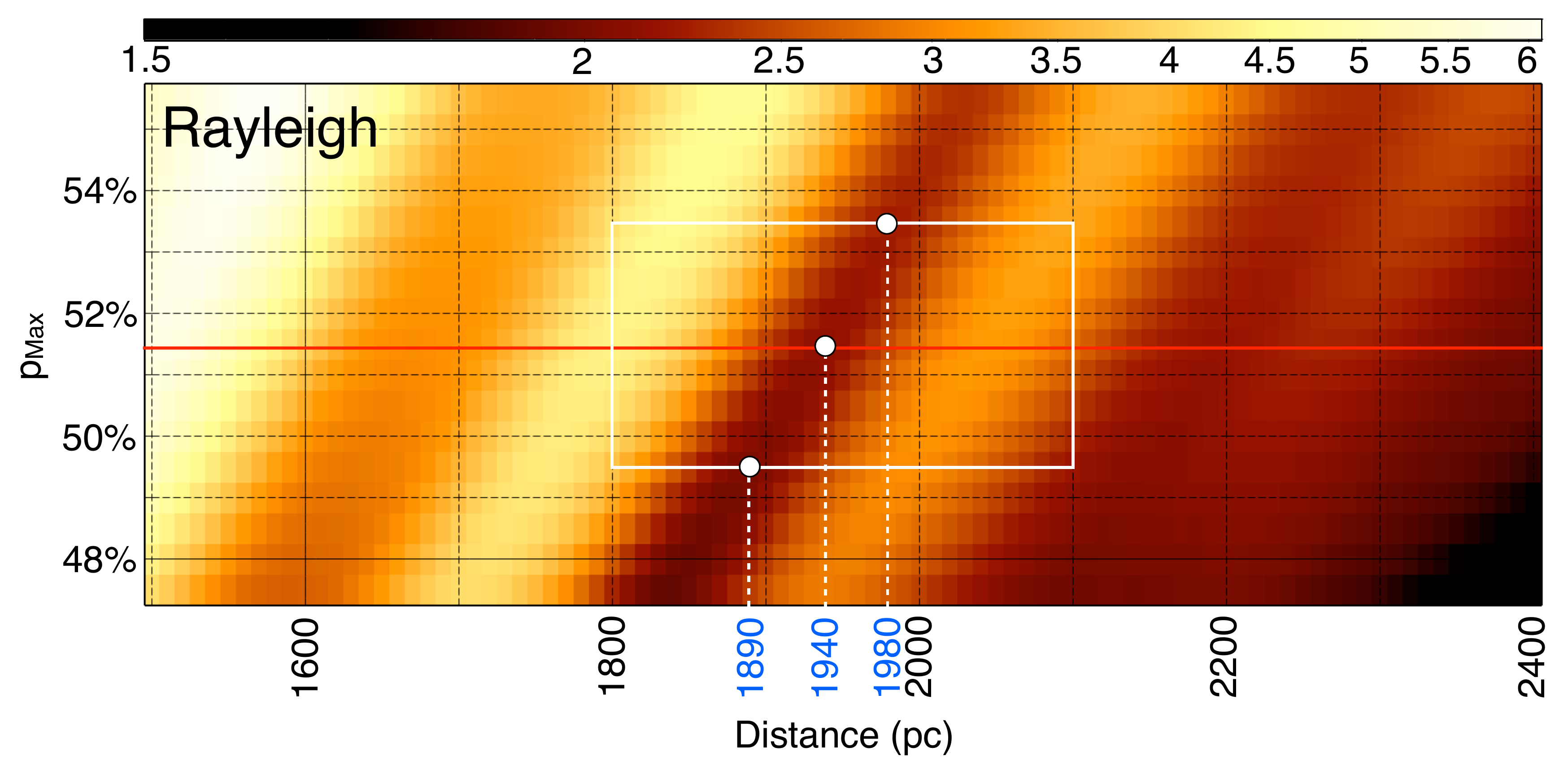} 
\caption{Map of the reduced $\chi^2$ of RS~Pup's distance adjustment for the scaled Milky Way dust polarization model ({\it left panel}) and the scaled Rayleigh scattering polarization model with $\theta_\mathrm{max}=\thetamax^\circ$ ({\it right panel}), as a function of the adopted $p_\mathrm{max}$ value (vertical axis). The white rectangle indicates horizontally the distance range found in the literature and vertically the adopted $p_\mathrm{max}$ uncertainty range ($\pmax\% \pm \pmaxerr\%$). The white dots show the position of the minimum $\chi^2$ for $p_\mathrm{max} = 49.5$, $\pmax$ and 53.5\%. The color scale shown above the map is a function of the square root of the reduced $\chi^2$ value. \label{chi2map}}
\end{figure*}

\subsubsection{Distance from the Milky Way dust polarization model}

The scaled MW dust model that we adopt has a single parameter $p_\mathrm{max} = \pmax\% \pm \pmaxerr\%$ (Sect.~\ref{drainemodel}). Fig.~\ref{chi2map} (left panel) shows the map of the reduced $\chi^2$ of the distance adjustment using this model. We estimated the statistical uncertainty by bootstrapping the distribution of the best-fit distances obtained on each selected pixel for the $p_\mathrm{max} = \pmax\%$ model. Thanks to the relatively large number of pixels, it is negligible ($< 10$\,pc) compared to the systematic uncertainties.

The $p_\mathrm{max}$ uncertainty domain and the distance range from the literature are represented as a rectangular box in Fig.~\ref{chi2map}. The lowest $\chi^2$ region is elongated and tilted with respect to the two axes, due to the correlation between $p_\mathrm{max}$ and the distance. Over the adopted range of $p_\mathrm{max}$ values, the minimum $\chi^2$ is reached between 1850 and 1970\,pc, i.e.~$\pm 60$\,pc around a best-fit distance of \distdraine\,pc for $p_\mathrm{max} = \pmax\%$.
Formally, our MW dust model has only one parameter $p_\mathrm{max}$, but there is also an implicit systematic uncertainty on the angle of maximum scattering, computed by \citetads{2003ApJ...598.1017D} using a specific dust mixture. As discussed in Sect.~\ref{rayleighmodel}, we adopt a range of $\pm 2^\circ$ around the maximum polarization angle of $\thetamax^\circ$, which translates into a systematic uncertainty of $\pm 2.5\%$ on the distance (Sect.~\ref{distanceRayleigh}). This converts to a systematic uncertainty of $\pm 50$\,pc on the distance.

We conclude that the best-fit distance for RS~Pup considering the scaled MW dust polarization model is $d(\mathrm{MW}) = \distdraine \pm 10_\mathrm{stat} \pm 60_\mathrm{pmax} \pm 50_{\theta \mathrm{max}}$\,pc. Combining all uncertainties, we therefore obtain $d(\mathrm{MW}) = \distdraine \pm \distdraineerr$\,pc (\distdraineerrpercent).

\subsubsection{Distance from the Rayleigh polarization model \label{distanceRayleigh}}
 
The overall structure of the $\chi^2$ map for the Rayleigh polarization model (Fig.~\ref{chi2map}, right panel) shows minor differences with the one obtained for the MW dust model. The tilt of the minimum $\chi^2$ regions with respect to the axes is slightly different, and the width of the minimum $\chi^2$ ``valley'' is narrower. These two effects are due to the more peaked polarization curve as a function of the scattering angle (see Fig.~\ref{polarizationModel}), which makes the scattering angle less correlated with the value of $p_\mathrm{max}$. The best fit distance for $p_\mathrm{max} = \pmax\%$ and $\theta_\mathrm{max}=\thetamax^\circ$ is \distrayleigh\,pc. The range of distances corresponding to $p_\mathrm{max}$ varying from \pmaxmin\% to \pmaxmax\% is 1890 to 1980\,pc, i.e.~$\pm 50$\,pc.

According to \citetads{2008AJ....135..605S}, for an instantaneous outburst, the dependence of the distance $d$ with the angle of maximum polarization $\theta_\mathrm{max}$ follows the relation:
\begin{equation}
d = d_\mathrm{90^\circ} \frac{1 + \cos \theta_\mathrm{max} }{\sin \theta_\mathrm{max}}
\end{equation}
where $d_\mathrm{90^\circ}$ is the distance obtained for $\theta_\mathrm{max} = 90^\circ$. From this expression, and considering the adopted range of $\theta_\mathrm{max}$ values ($\thetamax^\circ \pm 2^\circ$), we obtain a relative uncertainty of $\pm 3.5\%$ on the distance.
The case of RS~Pup is however slightly different, as we observe several occurrences of the light echoes simultaneously in the nebula, which we fit together using our model. The light curve of RS~Pup is also not perfectly sharp. We checked the uncertainty introduced by the $\theta_\mathrm{max}$ parameter by computing $\chi^2$ maps for the 1-$\sigma$ interval ($90-94^\circ$). From these maps, we obtain a relative uncertainty on the best-fit distance of $\pm 2.5\%$, corresponding to $\pm 50$\,pc, which we adopt as the systematic uncertainty associated with the maximum polarization angle (also for the distance estimated using the MW dust model).

The best-fit distance for RS~Pup with the Rayleigh dust polarization model is therefore $d(\mathrm{Rayleigh}) = \distrayleigh \pm \distrayleigherr$\,pc (\distrayleigherrpercent).

\subsection{Discussion and final distance \label{distancediscussion}}

We obtain distances of $d=\distdraine \pm \distdraineerr$\,pc and $\distrayleigh \pm \distrayleigherr$\,pc for our two scattering models, respectively, in good agreement with each other. The Milky Way dust polarization model is based on a more realistic physical model, and it is also found to be applicable to the light echoes observed by \citetads{2008AJ....135..605S} around the red transient V838~Mon. We therefore prefer the MW model over the Rayleigh scattering model.

\begin{figure}[]
\centering
\includegraphics[width=\hsize]{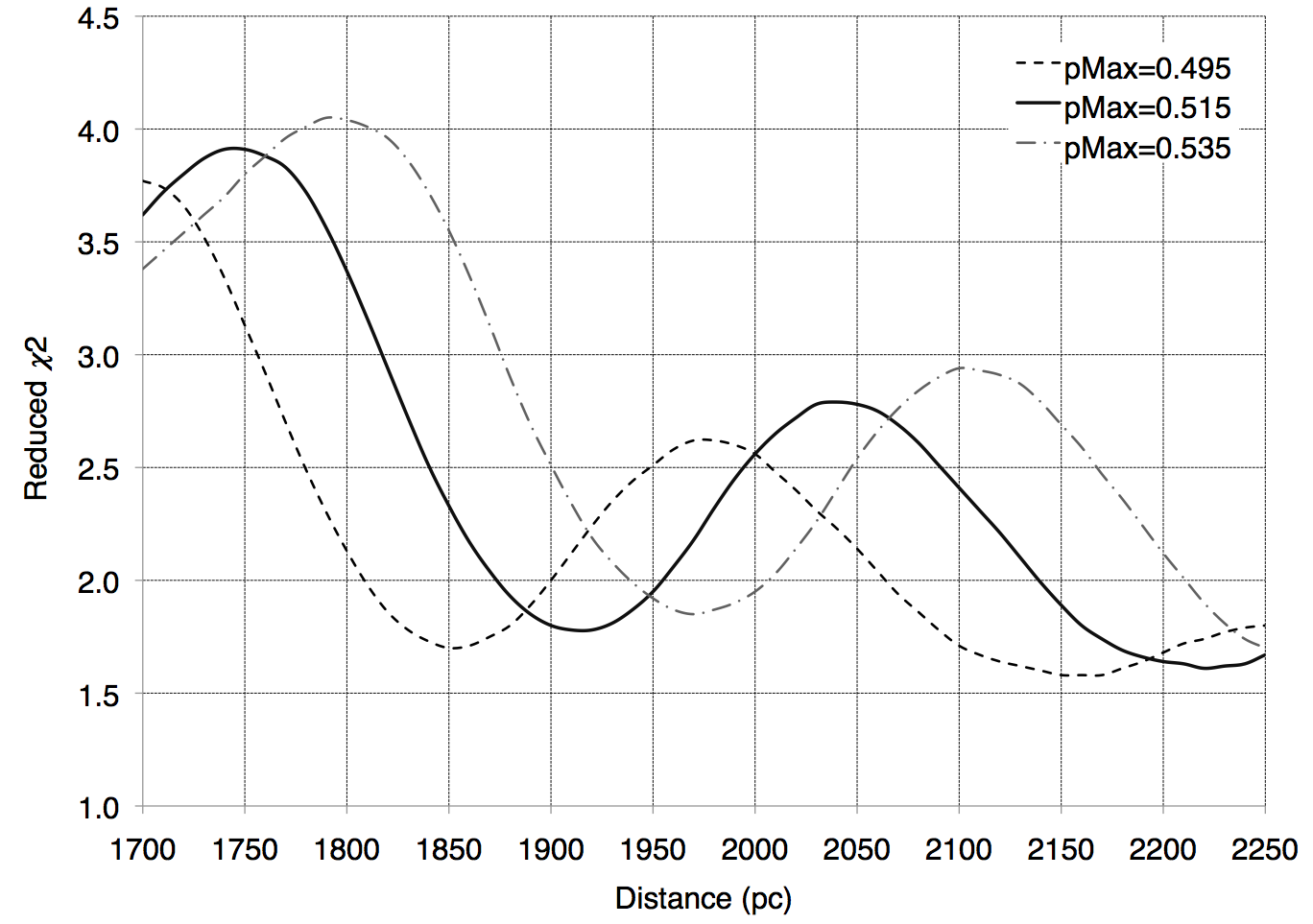} 
\caption{Reduced $\chi^2$ curve of the distance fit for the scaled Milky Way dust polarization model with $p_\mathrm{max}=\pmaxmin\%$, \pmax\% and \pmaxmax\%. \label{chi2cut}}
\end{figure}

A cut through the $\chi^2$ map for the MW dust model with $p_\mathrm{max} = \pmax\%$ and the extremes of the adopted uncertainty range (\pmaxmin\% and \pmaxmax\%) is presented in Fig.~\ref{chi2cut}. This curve presents a second minimum for $d = 2220$\,pc. This behavior is caused by the fact that the dust features suitable for the distance adjustment are located relatively far from RS~Pup (Fig.~\ref{IxPLmap}), between $30\arcsec$ and $1\arcmin$ from the Cepheid. This implies that the integral number of variation cycles $N$ between their photometric variation and that of the Cepheid is typically 6 (between 4 and 8). We therefore have an integral uncertainty of 1/6 of the distance, i.e.~$\approx 300$\,pc, due to the phase lag aliasing. While the shorter distance of $\approx 1600$\,pc can be excluded based on the $\chi^2$ criterion, the larger distance $\approx 2200$\,pc gives a fit of similar quality as \distdraine\,pc.

We discuss in Sect.~\ref{literaturedistances} why we believe that the larger distance can be excluded, and we propose a final distance value of $d=\distdraine \pm \distdraineerr$\,pc (\distdraineerrpercent) for RS~Pup.

\section{Discussion}

\subsection{Existing distance estimates \label{literaturedistances}}

A list of the published measurements of the distance of RS~Pup is presented in Table~\ref{distancesrspup}. The values derived from the Baade-Wesselink (BW) technique are listed together with the assumed value of the projection factor ($p$-factor). The $p$-factor is a multiplicative parameter used to convert radial velocities into pulsational velocities to determine distances through the BW technique. It is particularly important, as the distances derived from the BW technique scale linearly with the $p$-factor. This parameter has been measured only for two Cepheids: for $\delta$\,Cep ($P=5.37$\,d, $p=1.27 \pm 0.06$) by \citetads{2005A&A...438L...9M} and for OGLE-LMC-CEP-0227 ($P=3.80$\,d, $p=1.21 \pm 0.05$) by \citetads{2013MNRAS.436..953P}.

The measured BW distances in Table~\ref{distancesrspup} show both a statistical dispersion and a systematic dispersion caused by different choices of the $p$-factor value. As shown in Table~\ref{distancesrspup}, most authors based their BW distance determination on the linear period-$p$-factor relation established by \citetads{1986PASP...98..881H, 1989ApJ...341.1004H}: $p=1.39 - 0.03 \log P$, which gives $p = 1.341$ for RS~Pup. The distance of 1\,568\,pc published recently by \citetads{2013A&A...550A..70G} departs significantly from the other estimates, but this is due to the choice by this author of a particularly low value of the $p$-factor ($p = 1.112$). Scaling this estimate to the ``canonical'' $p=1.341$ results in a distance of 1\,891\,pc, in good agreement with our light echo measurement and the other values from the literature.
Excluding the \emph{Hipparcos} parallax values (which are unreliable for such a distant star) and the relatively low precision BW measurement by \citetads{1993ApJ...418..135G}, the weighted average of the measurements listed in Table~\ref{distancesrspup} results in a distance of 1960\,pc, with a standard deviation of 125\,pc. For homogeneity, we scaled all the BW distances using a $p$-factor of 1.341 for this averaging. As proposed by \citetads{2014A&A...566L..10A}, the cycle-to-cycle modulation of the radial velocity amplitude of RS~Pup probably contributes to the scattering of the BW distances in the literature. It should be noted that the listed measurements are not fully independent from each other, as they often rely on the same sets of observations (photometry, radial velocimetry).

So the overall agreement of the existing BW distance estimates with the value we derive in the present work is good within their uncertainties. This indicates that no significant bias is present on the distances determined with the BW technique for long-period Cepheids. We will discuss the implications of the distance we derive for RS~Pup on its $p$-factor in Paper~IV.

As discussed in Sect.~\ref{distancediscussion}, a minimum is present in the $\chi^2$ maps of our light echo model fit (Fig.~\ref{chi2map}) for a distance of $\approx 2200$\,pc (Fig.~\ref{chi2cut}). Such a large distance for RS~Pup appears unlikely as it would result in an unrealistically high value for the $p$-factor. Scaling the average distance of 1960\,pc from the literature ($p = 1.341$) to a distance of 2200\,pc would result in a $p$-factor of 1.51. Such a value is larger than the purely geometrical component of the $p$-factor (1.5), i.e.~the straight integration of the velocity field over a sphere of uniform brightness \citepads{2007A&A...471..661N}. A $p$-factor higher than 1.5 would imply that the limb of the star is brighter than the center of its apparent disk. While this cannot be formally excluded, it is not foreseen by existing models, in particular for long-period pulsating stars as RS~Pup \citepads[see e.g. the references in][]{2014IAUS..301..145N}. In addition, a distance of 2200\,pc would result in a significant deviation of RS~Pup from the period-luminosity relations (Sect.~\ref{PLrelation}) by $\approx 0.25$\,mag. Such a shift would exceed significantly the intrinsic dispersion of the P-L relation, particularly at infrared wavelengths \citepads[$\sigma(K) \approx 0.11$]{2012ApJ...744..132M}, and make RS~Pup a uniquely bright Cepheid, which appears unlikely.


\begin{table}
\caption{Observationally determined distances of RS~Pup (``IRSB" stands for ``infrared surface brightness").}
\label{distancesrspup}
\begin{tabular}{llll}
\hline
Distance [pc] & Method & $p$-factor & Source \\
\hline
\noalign{\smallskip}
$1\,780 \pm200$ & Light echoes & $-$ & H72 \\
\noalign{\smallskip}
$2\,580 \pm 300$ & IRSB & 1.341 & G93 \\
\noalign{\smallskip}
2\,040 & Hipparcos & $-$ & P97 \\
\noalign{\smallskip}
1\,930 & ZAMS & $-$ & DB97 \\
\noalign{\smallskip}
$2\,111^{+75}_{-73}$ & IRSB & 1.341 & F03 \\
\noalign{\smallskip}
$2\,052^{+61}_{-59}$ & IRSB & 1.341 & S04 \\
\noalign{\smallskip}
$1\,800 \pm 130$ & IRSB & 1.341 & T03 \\
\noalign{\smallskip}
$1\,706 \pm 228$ & IRSB (Bayesian) & 1.341 & B03 \\
\noalign{\smallskip}
$2\,004\pm 59$ & IRSB (least squares) & 1.341 & B05 \\
\noalign{\smallskip}
$1\,984\pm 122$ & IRSB (Bayesian) & 1.341 & B05 \\
\noalign{\smallskip}
$521^{+266}_{-132}$ & Hipparcos & $-$ & VL07a \\
\noalign{\smallskip}
$694^{+381}_{-181}$ & Hipparcos & $-$ & VL07b \\
\noalign{\smallskip}
$1\,830^{+109}_{-94}$ & IRSB & 1.341 & F07 \\
\noalign{\smallskip}
$1\,765 \pm 34 \pm 65$ & BW & 1.272  & G08 \\
\noalign{\smallskip}
$1\,810 \pm 30$ & IRSB & 1.249 & S11 \\
\noalign{\smallskip}
$1\,568 \pm 30 \pm 72$ & BW & 1.112 & G13 \\
\noalign{\smallskip}
\hline
\noalign{\smallskip}
$\distdraine \pm 80$ & Echoes (MW dust) & $-$ & K14 \\
\noalign{\smallskip}
$\distrayleigh \pm 70$ & Echoes (Rayleigh) & $-$ & K14 \\
\noalign{\smallskip}
\hline
\end{tabular}
\tablefoot{
B03: \citetads{2003ApJ...592..539B};
B05: \citetads{2005ApJ...631..572B};
DB97: \citetads{1997ApJ...486...60D};
P97: \citetads{1997ESASP1200.....P}; 
F03: \citetads{2003LNP...635...21F};
F07: \citetads{2007A&A...476...73F};
G08: \citetads{2008A&A...488...25G};
G13: \citetads{2013A&A...550A..70G};
G93: \citetads{1993ApJ...418..135G};
H72: \citetads{1972A&A....16..252H};
K14: present work;
S04: \citetads{2004A&A...415..531S};
S11: \citetads{2011A&A...534A..94S};
T03: \citetads{2003A&A...404..423T};
VL07a: \citetads{2007A&A...474..653V};
VL07b: \citetads{2007MNRAS.379..723V}.
}
\end{table}

\subsection{RS~Pup and the Period-Luminosity relation \label{PLrelation}}

\begin{figure}[]
\centering
\includegraphics[width=\hsize]{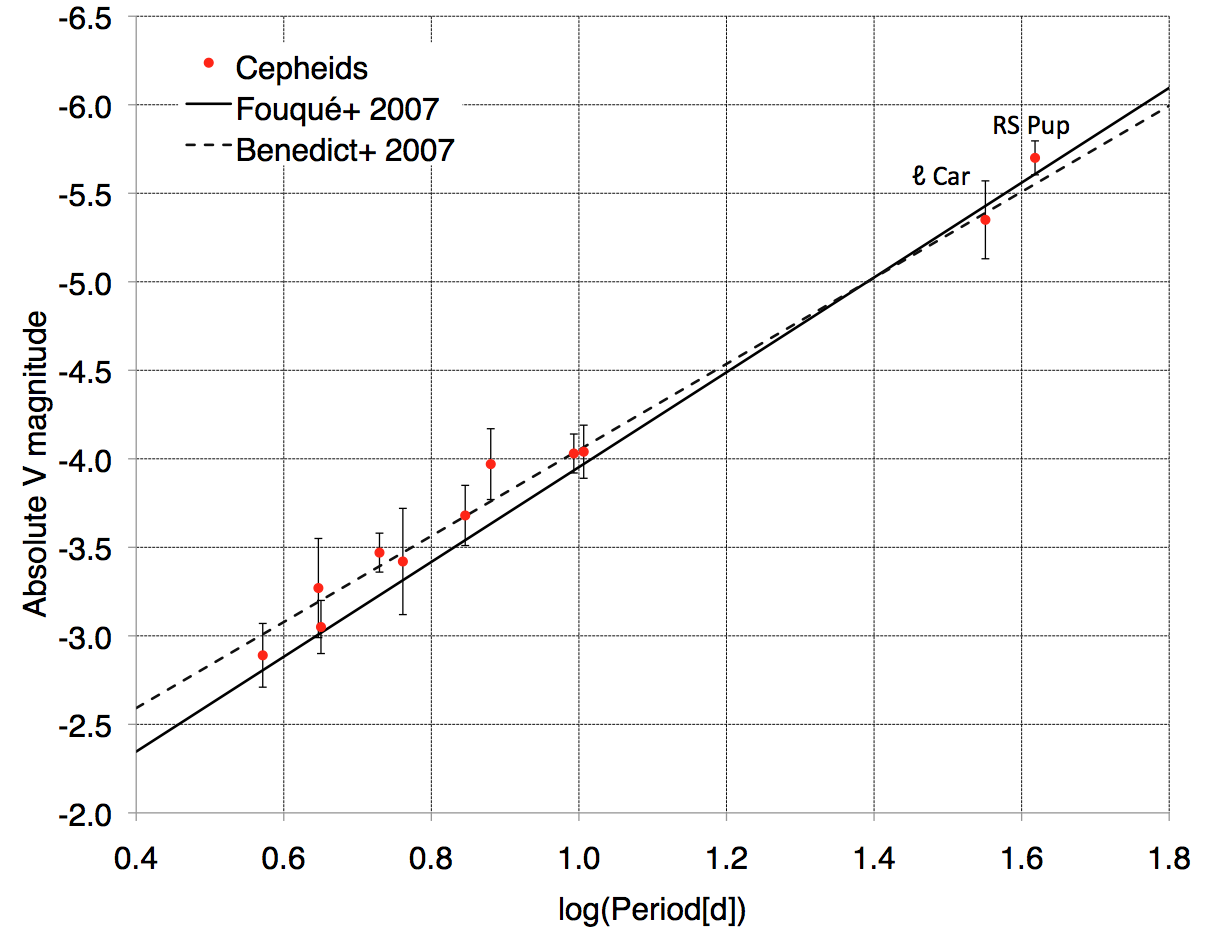} 
\caption{Period-luminosity diagram in the $V$ band showing the position of RS~Pup. The relations by \citetads{2007A&A...476...73F} and \citetads{2007AJ....133.1810B} are shown as solid and dashed lines, respectively.\label{PLrelationV}}
\end{figure}

Considering the intensity-mean magnitudes of $\langle V \rangle = 7.119$, $\langle B \rangle = 8.645$, and $\langle B-V \rangle = 1.528$ determined from our recent {\em SMARTS} photometry (Paper~IV) and using the values of $E(B-V) = 0.457 \pm 0.009$ \citepads{2007A&A...476...73F}, $R_V = 3.1$, and the distance of $\distdraine \pm \distdraineerr$\,pc, the absolute magnitudes of RS~Pup turn out to be $\langle M_V \rangle = -5.70 \pm 0.09$ and $\langle M_B \rangle = -4.63 \pm 0.09$.
%
%
Fig.~\ref{PLrelationV} shows the position of RS~Pup in the $V$ band period-luminosity diagram, together with the sample of Galactic Cepheids for which trigonometric parallaxes were obtained by \citetads{2007AJ....133.1810B} using the \emph{HST} Fine Guidance Sensor (FGS). The absolute magnitudes in this plot have been taken from these authors. In this diagram the linear PL relation derived by \citetads{2007AJ....133.1810B} from the FGS sample and the one from \citetads{2007A&A...476...73F} are shown for comparison. The agreement of RS\,Pup's position with both calibrations of the PL relation is satisfactory at the $\approx 1\sigma$ level, although the slope of the relation from \citetads{2007AJ....133.1810B} appears slightly too shallow. We reserve a more detailed discussion of this point to a forthcoming article (Paper~IV).

\section{Conclusion}

The combination of ACS imaging polarimetry and photometry of the circumstellar nebula of RS~Pup allowed us to measure the distance of this fiducial long-period Cepheid, which we find to be $d = \distdraine \pm \distdraineerr$\,pc (\distdraineerrpercent) corresponding to a parallax $\pi = \parallax \pm \parallaxerr$\,mas. The major part of the associated uncertainty comes from the adopted polarization model for the light scattered off the nebular dust. Consistent results are obtained for scaled versions of a Rayleigh scattering and a Milky Way dust model. The derived distance is compatible with existing Baade-Wesselink (BW) estimates form the literature, assuming a projection factor $p \approx 1.3$ (see Paper~IV for a detailed discussion).
This agreement between fully independent techniques gives confidence that the current calibration of the Cepheid distance scale is not affected by a large bias, as it is mainly based on the application of various implementations of the BW technique.
As discussed by \citetads{2014arXiv1406.1718B}, reaching a 1\% concensus on the value of the Hubble constant $H_0$ from high and low redshift indicators requires an unbiased calibration of the Cepheid distance scale. This can only be obtained from multiple and independent techniques, and in this context, our determination of the distance of RS~Pup from its remarkable light echoes brings an original contribution.

Considering the angular scale of the light echoes of RS~Pup in the plane of the sky ($\approx 4\arcsec$ at 1.9\,kpc), the measurement technique we applied is in principle applicable up to much larger distances. The angular resolution of the \emph{HST} would in principle allow distance measurements up to 20\,kpc, possibly even up to the Magellanic Clouds, for a configuration similar to RS~Pup. Unfortunately, this Cepheid may be the only star of its class in the Milky Way to be so closely embedded in a dusty reflection nebula\footnote{A ground-based imaging survey by one of us (H.E.B., in preparation) of $\sim$100 Galactic Cepheids failed to reveal any new associated reflection nebulae.}. As discussed by \citetads{2012A&A...541A..18K}, the bulk of the material surrounding RS~Pup is not the result of mass loss from the star itself, but consists of an interstellar dust cloud in which the Cepheid is temporarily embedded.
Such an association of a long-period Cepheid and a nebula is therefore the result of the unlikely conjunction of a short-lived massive star \citepads[$M \approx 9.2 \pm 0.5\,M_\odot$, from][]{2014A&A...564A.100A} in a brief phase of its evolution into the classical instability strip, and a dense interstellar dust cloud with a suitable geometry.
The shape of the RS~Pup nebula determined from polarimetry indicates that the high luminosity of the Cepheid and its progenitor affected the dust distribution, carving a large cavity in the dust cloud. The Cepheid may therefore clear the remaining dust over the next few million years, making the light echoes disappear.
It should be noted however that extended emission and infrared excess have been identified by \citetads{2011AJ....141...42B} and \citetads{2010ApJ...725.2392M} around several Cepheids from \emph{Spitzer} observations. The short-period Cepheid \object{SU~Cas} \citepads[see e.g.~][]{2012MNRAS.422.2501T} is also located close to a light-scattering dust cloud. But the unfavorable geometry of the cloud, the short period of the star (2\,days) and the low amplitude of its photometric variation ($\Delta m_V \approx 0.4$\,mag) will probably make the detection of the echoes difficult.

The Gaia astrometric satellite, successfully launched in December 2013, will provide within a few years a test of our distance estimate \citepads[see e.g.\,][]{2011A&A...530A..76W}. RS~Pup would also be an excellent candidate for a trigonometric parallax measurement using the newly commissioned spatial scanning technique with the \emph{HST} Wide Field Camera 3 \citepads{2014ApJ...785..161R}, which has the potential to provide a comparable accuracy to our measurement .

\begin{acknowledgements}
We thank Dr~Daniel~Rouan for fruitful discussions on the determination of the geometry of RS~Pup's nebula.
We acknowledge financial support from the ``Programme National de Physique Stellaire" (PNPS) of CNRS/INSU, France.
Support for Program number GO-13454 was provided by NASA through a grant from the Space Telescope Science Institute, which is
operated by the Association of Universities for Research in Astronomy, Incorporated, under NASA contract NAS5-26555.
PK and AG acknowledge support of the French-Chilean exchange program ECOS-Sud/CONICYT.
LSz acknowledges support from the ESTEC Contract No.\,4000106398/12/NL/KML.
AG acknowledges support from FONDECYT grant 3130361.
This research received the support of PHASE, the high angular resolution partnership between ONERA, Observatoire de Paris, CNRS and University Denis Diderot Paris 7.
We used the SIMBAD and VIZIER databases at the CDS, Strasbourg (France), and NASA's Astrophysics Data System Bibliographic Services.
We used the IRAF package, distributed by the NOAO, which are operated by the Association of Universities for Research in Astronomy, Inc., under cooperative agreement with the National Science Foundation.
Some of the data presented in this paper were obtained from the Multimission Archive at the Space Telescope Science Institute (MAST). STScI is operated by the Association of Universities for Research in Astronomy, Inc., under NASA contract NAS5-26555. Support for MAST for non-HST data is provided by the NASA Office of Space Science via grant NAG5-7584 and by other grants and contracts.
\end{acknowledgements}

\bibliographystyle{aa} 
\bibliography{biblioRSPup}

\Online{
\begin{appendix}
\section{Computation of the polarimetric quantities\label{appendixpolar}}

The Stokes parameters $I$, $Q$ and $U$ were derived through the relations \citepads{2008AJ....135..605S}:
\begin{equation}
\label{EqI}
I = \frac{2}{3} \left( \mathrm{POL0} + \mathrm{POL60} + \mathrm{POL120} \right)
\end{equation}
\begin{equation}
\label{EqQ}
Q = \frac{2}{3} \left( 2\,\mathrm{POL0} - \mathrm{POL60} - \mathrm{POL120} \right)
\end{equation}
\begin{equation}
\label{EqU}
U = \frac{2}{\sqrt{3}} \left( \mathrm{POL60} - \mathrm{POL120} \right)
\end{equation}
where POL0, POL60 and POL120 are the images of RS~Pup in the three polarizers, after the PSF wings and sky background were subtracted. The degree of linear polarization $p_L$ was then derived through:
\begin{equation}
\label{EqpL}
p_L = \frac{\sqrt{Q^2 + U^2}}{I} - B(\mathrm{SNR})
\end{equation}
where $B(\mathrm{SNR})$ is the bias correction depending on the photometric signal-to-noise ratio (SNR) as listed in Table~A.1 of  \citetads{1999PASP..111.1298S}.
The polarization electric-vector position angle in detector coordinates $\theta_D$ was derived through the expression:
\begin{equation}
\label{EqTheta}
\theta_D = \frac{1}{2} \tan^{-1} \left( \frac{U}{Q} \right) - 38.2^\circ.
\end{equation}

The uncertainties were propagated to the Stokes parameters starting from the photometric error bars produced by the ACS processing pipeline
  $\sigma(\mathrm{POL0})$, $\sigma(\mathrm{POL60})$ and $\sigma(\mathrm{POL120})$ through:
\begin{equation}
\label{EqErrI}
\sigma^2(I) = \left( 2/3 \right)^2 \left[ \sigma^2(\mathrm{POL0}) + \sigma^2(\mathrm{POL60}) + \sigma^2(\mathrm{POL120}) \right]
\end{equation}
\begin{equation}
\label{EqErrQ}
\sigma^2(Q) = \left( 2/3 \right)^2 \left[ 4\,\sigma^2(\mathrm{POL0}) + \sigma^2(\mathrm{POL60}) + \sigma^2(\mathrm{POL120}) \right]
\end{equation}
\begin{equation}
\label{EqErrU}
\sigma^2(U) = \left(4/3 \right) \left[ \sigma^2(\mathrm{POL60}) + \sigma^2(\mathrm{POL120}) \right]
\end{equation}
and to the degree of linear polarization $p_L$ through:
\begin{equation}
\label{EqVarUI}
\sigma^2(U/I) = \frac{U^2}{I^2} \ \left( \frac{\sigma^2(U)}{U^2} + \frac{\sigma^2(I)}{I^2} - \frac{\sigma^4(I)}{I^4} \right)
\end{equation}
\begin{equation}
\label{EqVarQI}
\sigma^2(Q/I) = \frac{Q^2}{I^2} \ \left( \frac{\sigma^2(Q)}{Q^2} + \frac{\sigma^2(I)}{I^2} - \frac{\sigma^4(I)}{I^4} \right)
\end{equation}
\begin{equation}
\label{EqVarPL}
\sigma^2(p_L) = \frac{\sigma^2(U)\ \sigma^2(U/I) + \sigma^2(Q)\  \sigma^2(Q/I)}{(Q + U)^2} 
\end{equation}
%

\end{appendix}
} 

\end{document}